# Two-parameter classes of exactly solvable quantum systems

A. D. Alhaidari

*Saudi Center for Theoretical Physics, P.O. Box 32741, Jeddah 21438, Saudi Arabia*

**Abstract:** We introduce two-parameter classes of exactly-solvable novel systems whose Hamiltonian operators could be represented by tridiagonal symmetric matrices in some orthogonal bases. The associated wavefunction is written as point-wise convergent series in the basis elements. The expansion coefficients of the series are orthogonal polynomials in the energy that satisfy the resulting three-term recursion relation starting with two-parameter initial values. These polynomials contain all physical information about the system and they depend on the values of the two parameters. We obtain the associated two-parameter potential function induced by the change in the initial values that causes the system's wavefunction to change. We give several illustrative examples of these systems with continuous and/or discrete energy spectra. Moreover, a curious phenomenon is observed where bound states and/or resonances are induced in a system with pure continuous spectrum (e.g., a free particle) if the two parameters in the initial values exceed certain critical limits.

## 1. Introduction

In quantum mechanics, two objects are essential for the description of a physical system. The wavefunction $\chi(t,\vec{r})$, which gives all physical information about the system at any given time, and the Hamiltonian operator $H$ that provides for the temporal development of the system. These two objects are linked by the linear Schrödinger wave equation $\mathrm{i}\partial_t\chi = H\chi$ (in the atomic units $\hbar = m = 1$) subject to some boundary conditions. For time-independent Hamiltonians, the time separates in the wavefunction making $\chi(t,\vec{r}) = e^{-\mathrm{i}Et}\psi(E,\vec{r})$ and the Schrödinger wave equation becomes $H\psi = E\psi$ where $E$ is the energy of the system. The potential function, on the other hand, is a classical concept that does not appear in the postulates of quantum mechanics. Writing the Hamiltonian as the sum of the kinetic energy operator and a potential function is a choice that was carried over from classical mechanics. In fact, countless quantum systems, especially in condensed matter physics, are treated in the literature without reference to, or explicit construction of, any potential function. In such cases, it is sufficient that one either write down explicitly the wavefunction $\chi(t,\vec{r})$ or give some representation of the Hamiltonian operator together with the boundary conditions. In this study, we work in one-dimension where $\chi(t,x) = e^{-\mathrm{i}Et}\psi(E,x)$ and $x_- \le x \le x_+$ with $x_\pm$ being the boundaries of space that could be finite or infinite. Moreover, we write the Hamiltonian as $H = H_0 + V$ where $H_0$ is the reference Hamiltonian operator and $V$ is the interaction potential.

The energy spectrum of a physical system is either pure continuous (like that of a free particle) or pure discrete (like that of the harmonic oscillator) or a combination thereof (like that of a particle in finitely deep well). The continuous and discrete energy spectra do not overlap. Moreover, the continuous spectrum consists generally of one or more continuous but disconnected intervals (called energy bands). However, the discrete spectrum can contain finite or countably infinite number of discrete points. The total wavefunction of the systems can thus be written as a Fourier expansion over the entire energy spectrum:

$$\chi(t,x)=\int_{\Omega} e^{-\mathrm{i}Et}\psi(E,x)dE+\sum_{j=0}^{N} e^{-\mathrm{i}E_j t}\psi_j(x)\,, \tag{1}$$

where $\Omega$ is the continuous energy interval(s) (typically, $E\geq 0$) and $N$ could be finite or infinite (typically, $E_j<0$). We write the Fourier components of the wavefunction as the following infinite, but point-wise convergent, series

$$\psi(E,x)=\sum_{n=0}^{\infty} f_n(E)\phi_n(\lambda x)=f_0(E)\sum_{n=0}^{\infty} P_n(z)\phi_n(\lambda x)\,, \tag{2a}$$

$$\psi_j(x)=\sum_{n=0}^{\infty} g_n(E_j)\phi_n(\lambda x)=g_0(E_j)\sum_{n=0}^{\infty} P_n(z_j)\phi_n(\lambda x)\,, \tag{2b}$$

where $z$ is called the *spectral parameter* which is a real function of the energy and $\lambda$ is a scale parameter of inverse length dimension. $\{P_n(z)\}_{n=0}^{\infty}$ is a set of real functions with $P_0(z)=1$ since we wrote $f_n=f_0P_n$ and $g_n=g_0P_n$. $\{\phi_n(\lambda x)\}_{n=0}^{\infty}$ is a complete set of functions in configuration space that form an invariant subspace in the domain of the reference Hamiltonian (i.e., $H_0\,\phi_n\in\{\phi_m\}_{m=0}^{\infty}$). Specifically, we require that

$$H_0\left|\phi_n(\lambda x)\right\rangle=\tfrac{1}{2}\lambda^2\left[a_n^{\mu}\left|\phi_n(\lambda x)\right\rangle+b_n^{\mu}\left|\phi_{n+1}(\lambda x)\right\rangle+b_{n-1}^{\mu}\left|\phi_{n-1}(\lambda x)\right\rangle\right], \tag{3}$$

for $n=1,2,3,...$ with $\mu$ being a kinetic parameter or set of parameters, and $\{a_n^{\mu},b_n^{\mu}\}$ are real dimensionless coefficients such that $b_n^{\mu}\neq 0$ for all $n$ and by definition $b_{-1}^{\mu}:=0$ (Hence, $H_0\left|\phi_0\right\rangle=\frac{1}{2}\lambda^2[a_0^{\mu}\left|\phi_0\right\rangle+b_0^{\mu}\left|\phi_1\right\rangle]$). We start by solving the reference problem (i.e., taking $V=0$) where the wave equation becomes $H_0\,\chi=E\chi$, giving

$$\left(2E/\lambda^2\right)p_n(z)=a_n^{\mu}p_n(z)+b_n^{\mu}p_{n+1}(z)+b_{n-1}^{\mu}p_{n-1}(z)\,, \tag{4a}$$

$$\left(2E_j/\lambda^2\right)p_n(z_j)=a_n^{\mu}p_n(z_j)+b_n^{\mu}p_{n+1}(z_j)+b_{n-1}^{\mu}p_{n-1}(z_j)\,, \tag{4b}$$

for $n=1,2,3,...$ and we have assumed that the basis elements are orthonormal (i.e., $\left\langle\phi_n|\phi_m\right\rangle=\delta_{n,m}$). Equation (4) is a symmetric three-term recursion relation that gives $P_n(z)$ as a polynomial in $z=2E/\lambda^2$ of degree $n$ (called the *spectral polynomial*) with the initial values $P_0(z)=1$ and $P_1(z)=\alpha z-\beta$ where $\alpha$ and $\beta$ are arbitrary dimensionless parameters independent of the energy (also, independent of $z$) such that $\alpha\neq 0$. The spectral polynomials depend on these two parameters since the initial values $P_0(z)$ and $P_1(z)$ will propagate throughout the recursion (4) [1,2]. That is, $\{P_n(z)\}_{n=0}^{\infty}$ changes with these parameters and so does the wavefunction (2). Therefore, the system is parameterized by $\{\mu,\alpha,\beta\}$ and its wave-function components read

$$\psi_{\mu}^{(\alpha,\beta)}(E,x)=f_0(E)\sum_{n=0}^{\infty} P_n^{(\alpha,\beta)}(z)\phi_n^{\mu}(\lambda x)\,. \tag{5a}$$

$$\psi_{j,\mu}^{(\alpha,\beta)}(x) = g_0(E_j)\sum_{n=0}^{\infty} P_n^{(\alpha,\beta)}(z_j)\phi_n^{\mu}(\lambda x)\,. \tag{5b}$$

The energy factors $f_0(E)$ and $g_0(E_j)$ can be determined by normalization (for the discrete bound states) and boundary conditions (for the continuum scattering states). The generalized orthogonality of the spectral polynomials reads as follows [3,4]:

$$\int_{\Omega} \rho(z) P_n^{(\alpha,\beta)}(z) P_m^{(\alpha,\beta)}(z)\,dz + \sum_{j=0}^{N} \omega_j P_n^{(\alpha,\beta)}(z_j) P_m^{(\alpha,\beta)}(z_j) = \delta_{n,m}\,, \tag{6}$$

where $\rho(z)$ is the continuous weight function and $\omega_j$ is the discrete weight function. Both are positive definite. For systems with pure continuous/discrete energy spectrum, the sum/integral in the orthogonality (6) and total wavefunction (1) is absent.

The wavefunction components (5) are solutions of the reference problem ($V=0$) if the recursion relation (4) is also satisfied for $n=0$. That is, if $\alpha = 1/b_0^{\mu} := \hat{\alpha}$ and $\beta = a_0^{\mu}/b_0^{\mu} := \hat{\beta}$. For other values of $\alpha$ and/or $\beta$, the solution changes implying the influence of a potential $V \neq 0$ added to the reference Hamiltonian as $H = H_0 + V$. Thus, the potential is parameterized as $V_{\mu}^{(\alpha,\beta)}$ such that $V_{\mu}^{(\hat{\alpha},\hat{\beta})} = 0$. In section 2, we give a demonstration in a very simple system: the free particle in 1D, and show how an interaction potential is induced simply by changing the initial values of the associated spectral polynomials (i.e., taking $\alpha \neq \hat{\alpha}$ and/or $\beta \neq \hat{\beta}$). In sections 3, 4, and 5, we give several nontrivial examples and plot sample wavefunctions with $V = V_{\mu}^{(\hat{\alpha},\hat{\beta})} = 0$ and $V = V_{\mu}^{(\alpha,\beta)} \neq 0$. In section 6, we show that it is possible for the induced potential $V_{\mu}^{(\alpha,\beta)}$ in a system with pure continuous spectrum to generate bound states and/or resonances. In section 7, we show how to construct the potential function $V_{\mu}^{(\alpha,\beta)}(x)$ in configuration space. Finally, we conclude in section 8.

## 2. The 1D free particle class

As a simple illustration, we consider the free particle in 1D where $H_0 = -\frac{1}{2}\frac{d}{dx^2}$ and the reference wavefunction is just $\psi(E,x) = e^{ikx}$ with $k^2 = 2E$ and $x_{\pm} = \pm\infty$. Using Mehler's formula (See problem 23 on page 380 of Ref. [3]), the oscillatory function $e^{ikx}$ has the following point-wise convergent expansion

$$e^{ikx} = e^{i(k/\lambda)(\lambda x)} = \sqrt{2}\,e^{-k^2/2\lambda^2} e^{-\lambda^2 x^2/2} \sum_{n=0}^{\infty} \frac{i^n}{2^n n!} H_n(k/\lambda) H_n(\lambda x)\,, \tag{7}$$

where $H_n(y)$ is the Hermite polynomial of degree *n* in *y*. If we consider the real part, then the wavefunction reads

$$\psi(E,x) = \cos(kx) = \sqrt{2}\,e^{-k^2/2\lambda^2} e^{-\lambda^2 x^2/2} \sum_{n=0}^{\infty} \frac{(-1)^n}{2^{2n}(2n)!} H_{2n}(k/\lambda) H_{2n}(\lambda x)\,. \tag{8}$$

The imaginary part can also be treated in a similar fashion (see below). It is easy to show that if we choose the following orthonormal basis elements in (8)

$$\phi_n(\lambda x)=\sqrt{\lambda/\sqrt{\pi}}\,e^{-\lambda^2x^2/2}\left[H_{2n}(\lambda x)\Big/\sqrt{2^{2n}(2n)!}\right], \tag{9}$$

then by comparing Eq. (8) to Eq. (5a) we obtain the following spectral polynomials and weight factor $f_0(E)$

$$P_n^{(\hat{\alpha},\hat{\beta})}(z)=(-1)^n H_{2n}(\sqrt{z})\Big/\sqrt{2^{2n}(2n)!}, \tag{10a}$$

$$f_0(E)=\sqrt{\tfrac{2}{\lambda}\sqrt{\pi}\;e^{-z/2}}, \tag{10b}$$

where $z=(k/\lambda)^2=2E/\lambda^2$ and, as expected, the system has a pure continuous energy spectrum with $z\geq 0$. Using the orthogonality of the Hermite polynomials, $\int_{-\infty}^{+\infty}e^{-y^2}H_n(y)H_m(y)dy=\sqrt{\pi}\,2^n n!\,\delta_{n,m}$, we can show that $\rho(z)=e^{-z}/\sqrt{\pi z}$ and thus

$$f_0(E)=\sqrt{\tfrac{2\pi}{\lambda}\sqrt{z}\,\rho(z)}. \tag{10c}$$

Moreover, using the differential equation of the Hermite polynomials, $H_n''(y)-2yH_n'(y)+2nH_n(y)=0$, and their recursion relation, $yH_n(y)=nH_{n-1}(y)+\frac{1}{2}H_{n+1}(y)$, one can easily show that Eq. (3) is satisfied with $a_n=2n+\frac{1}{2}$ and $b_n=-\sqrt{(n+1)\left(n+\frac{1}{2}\right)}$. Thus, the two parameters of the reference system are $\alpha=\hat{\alpha}=-\sqrt{2}$ and $\beta=\hat{\beta}=-1/\sqrt{2}$. Now, let us change the parameter values to $\alpha=-1.0$ and $\beta=-1.2$ leading to a change in the spectral polynomials and the wavefunction (see Figure 1). This change implies that an interaction must have been induced corresponding to the total Hamiltonian $H=-\frac{1}{2}\frac{d}{dx^2}+V^{(\alpha,\beta)}(x)$. In section 7, we derive this two-parameter potential function as given by Eq. (46).

The solid trace in Figure 1 is the reference wavefunction $\cos(kx)$ superimposed by the value of the truncated series on the right-hand side of Eq. (8) as dotted trace. It is evident that the two traces coincide perfectly, especially when the number of terms in the series becomes large enough. The dashed curve in the figure is the result of calculating the series (8) where the spectral polynomials (10a) are replaced by $P_n^{(\alpha,\beta)}(z)$ that satisfies the exact same recursion relation (4a) but with the initial value $P_1(z)=\alpha z-\beta$ not $P_1(z)=\hat{\alpha}z-\hat{\beta}$. Figure 2, shows several plots of the wavefunction series for different values of $\alpha$ and $\beta$. Figure 3, shows plots of the weight function $\rho(z)$ for different values of $\alpha$ and $\beta$. The solid trace corresponds to $\alpha=\hat{\alpha}$ and $\beta=\hat{\beta}$ where $\rho(z)=e^{-z}/\sqrt{\pi z}$. The calculation of these weight functions was performed as shown in the Appendix.

Figure 4 is a reproduction of Figure 1 but for the imaginary part of the wavefunction, $\psi(E,x)=\sin(kx)$, where

$$\phi_n(\lambda x)=\sqrt{\lambda/\sqrt{\pi}}\,e^{-\lambda^2x^2/2}\left[H_{2n+1}(\lambda x)\Big/\sqrt{2^{2n+1}(2n+1)!}\right], \tag{11a}$$

$$P_n^{(\hat{\alpha},\hat{\beta})}(z)=\frac{(-1)^n}{\sqrt{2z}}H_{2n+1}(\sqrt{z})\Big/\sqrt{2^{2n+1}(2n+1)!}\,, \tag{11b}$$

$$f_0(E)=2\sqrt{\tfrac{z}{\lambda}\sqrt{\pi}}\,e^{-z/2}\,, \tag{11c}$$

$$\rho(z)=2\sqrt{z/\pi}\,e^{-z}\,,\qquad a_n=2n+\tfrac{3}{2}\,,\qquad b_n=-\sqrt{(n+1)\left(n+\tfrac{3}{2}\right)}\,. \tag{11d}$$

Thus, $\hat{\alpha}=1/b_0=-\sqrt{2/3}$, $\hat{\beta}=a_0/b_0=-\sqrt{3/2}$ and $f_0(E)=\sqrt{\frac{2\pi}{\lambda}\sqrt{z}\rho(z)}$. Note that the factor $1/\sqrt{2z}$ was introduced in the definition of the spectral polynomial (11b) to make $P_0(z)=1$ and $P_n(z)$ a polynomial in $z$ of degree $n$. However, to maintain the overall value of the wave-function, this same factor is absorbed into the function $f_0(E)$. Note also that multiplying the spectral polynomial by any function of $z$ like $1/\sqrt{2z}$ will not affect the polynomial solution of the recursion relation (4) as long as the function does not depend on the index $n$ and that the multiplication results in a polynomial in $z$. Figure 5 is a reproduction of Figure 3 for the weight function $\rho(z)$ with different values of $\alpha$ and $\beta$.

## 3. The 3D free particle class

In three dimensions with spherical symmetry, we take the reference Hamiltonian as the radial component of the kinetic energy operator

$$H_0=-\frac{1}{2}\frac{d^2}{dr^2}+\frac{\ell(\ell+1)}{2r^2}\,, \tag{12}$$

where $\ell$ is the angular momentum quantum number, $\ell=0,1,2,...$, and $r\geq 0$. We can show that the following orthonormal basis elements satisfy Eq. (3)

$$\phi_n^\ell(\lambda r)=\sqrt{\frac{2\lambda\Gamma(n+1)}{\Gamma(n+\ell+\frac{3}{2})}}\,(\lambda r)^{\ell+1}\,e^{-\lambda^2r^2/2}L_n^{\ell+\frac{1}{2}}(\lambda^2r^2)\,, \tag{13}$$

where $L_n^\nu(y)$ is the Laguerre polynomial giving $a_n^\ell=2n+\ell+\frac{3}{2}$ and $b_n^\ell=\sqrt{(n+1)\left(n+\ell+\frac{3}{2}\right)}$. Consequently, $\hat{\alpha}=1/\sqrt{\ell+\frac{3}{2}}$, $\hat{\beta}=\sqrt{\ell+\frac{3}{2}}$ and the spectral polynomial that satisfies the recursion relation (4) with the initial values $P_0(z)=1$ and $P_1(z)=\hat{\alpha}z-\hat{\beta}$ is

$$P_n^{(\hat{\alpha},\hat{\beta})}(z)=(-1)^n\sqrt{\Gamma(n+1)\Big/\left(\ell+\tfrac{3}{2}\right)_n}\,L_n^{\ell+\frac{1}{2}}(z)\,, \tag{14}$$

where $(c)_n=\Gamma(n+c)/\Gamma(c)=c(c+1)(c+2)...(c+n-1)$ is the Pochhammer symbol (rising/upper factorial). Hence, the reference wavefunction becomes

$$\psi_\ell^{(\hat{\alpha},\hat{\beta})}(E,r)=f_0(E)\sum_{n=0}^{\infty}P_n^{(\hat{\alpha},\hat{\beta})}(z)\phi_n^\ell(\lambda r)\,. \tag{15}$$

To determine the energy weight factor $f_0(E)$, we can compare (15) to the well-known regular solution of the reference wave equation $H_0\,\psi_\ell^{(\hat{\alpha},\hat{\beta})}(E,r)=E\,\psi_\ell^{(\hat{\alpha},\hat{\beta})}(E,r)$ that reads $\psi_\ell^{(\hat{\alpha},\hat{\beta})}(E,r)$ $=\sqrt{kr}J_{\ell+\frac{1}{2}}(kr)$ with $J_\nu(y)$ being the Bessel function of the first kind [5]. Thus, we obtain

$$f_0(E)=\sqrt{2/\lambda\,\Gamma\left(\ell+\tfrac{3}{2}\right)}\;z^{\frac{\ell+1}{2}}e^{-z/2}, \tag{16}$$

where we have used the following representation of the Bessel function [4]

$$J_\nu(kr)=\frac{2(kr)^\nu}{\Gamma(\nu+1)}e^{-\lambda^2r^2/2}e^{-k^2/2\lambda^2}\sum_{n=0}^{\infty}\frac{(-1)^n n!}{(\nu+1)_n}L_n^\nu(k^2/\lambda^2)L_n^\nu(\lambda^2r^2)\,. \tag{17}$$

The orthogonality of the Laguerre polynomials gives $\rho(z)=z^{\ell+\frac{1}{2}}e^{-z}/\Gamma\left(\ell+\tfrac{3}{2}\right)$ making $f_0(E)=$ $\sqrt{2\sqrt{z}\rho(z)/\lambda}$, which has the same functional form as that in section 2.

Now, we change the initial value to $P_1(z)=\alpha z-\beta$ with $\alpha\neq\hat{\alpha}$ and/or $\beta\neq\hat{\beta}$. Figure 6 compares the reference solution $\sqrt{kr}J_{\ell+\frac{1}{2}}(kr)$ as solid trace to the series solution (15) as dotted trace and the solution $\psi_\ell^{(\alpha,\beta)}(E,r)$ for some values of $\alpha$ and $\beta$. Figure 7 shows a comparison of the weight function of the spectral polynomials for the reference problem (solid trace) with those that correspond to $\alpha\neq\hat{\alpha}$ and/or $\beta\neq\hat{\beta}$ (see the Appendix for this calculation). In section 7, we obtain the potential function $V_\ell^{(\alpha,\beta)}(r)$ that caused the deviation in the wavefunction $\psi_\ell^{(\alpha,\beta)}(E,r)$ from the reference solution $\psi_\ell^{(\hat{\alpha},\hat{\beta})}(E,r)=\sqrt{kr}J_{\ell+\frac{1}{2}}(kr)$. It is given by Eq. (44).

# 4. The isotropic oscillator class

In 3D with spherical symmetry, we take the reference Hamiltonian as the radial component of the isotropic oscillator:

$$H_0=-\frac{1}{2}\frac{d^2}{dr^2}+\frac{\ell(\ell+1)}{2r^2}+\frac{1}{2}\kappa^2r^2, \tag{18}$$

where $\kappa$ is the oscillator frequency. In this case, we expect that the associated spectral polynomial to have a pure discrete spectrum of infinite size. Out of the many hypergeometric orthogonal polynomials in the Askey-Wilson scheme [6], two have such property: The Meixner polynomial and the Charlier polynomial. Anyhow, if we adopt the orthonormal basis elements (13) and use the differential equation, recursion relation, and differential property of the Laguerre polynomials, $y\frac{d}{dy}L_n^\nu(y)=nL_n^\nu(y)-(n+\nu)L_{n-1}^\nu(y)$, then we can show that Eq. (3) is satisfied with the following recursion coefficients

$$a_n^\ell=\left(\tfrac{\kappa^2}{\lambda^4}+1\right)\left(2n+\ell+\tfrac{3}{2}\right), \tag{19a}$$

$$b_n^\ell=-\left(\tfrac{\kappa^2}{\lambda^4}-1\right)\sqrt{(n+1)\left(n+\ell+\tfrac{3}{2}\right)}\,. \tag{19b}$$

Thus, $\hat{\alpha}=-1\Big/\left(\frac{\kappa^2}{\lambda^4}-1\right)\sqrt{\ell+\frac{3}{2}}$ and $\hat{\beta}=-\frac{\kappa^2+\lambda^4}{\kappa^2-\lambda^4}\sqrt{\ell+\frac{3}{2}}$. The spectral polynomial that satisfies the recursion relation (4) with these recursion coefficients and initial values $P_0(z)=1$, $P_1(z)=\hat{\alpha}z-\hat{\beta}$ is the discrete Meixner polynomial whose normalized version reads (see Appendix A of [7] or section 9.10 of Ref. [6])

$$P_n^{(\hat{\alpha},\hat{\beta})}(z_j)=\sqrt{\frac{\left(\ell+\frac{3}{2}\right)_n}{n!}}\,\sigma^n\,{}_2F_1\left(\left.\begin{matrix}-n,-j\\ \ell+\frac{3}{2}\end{matrix}\right|1-\sigma^{-2}\right), \tag{20}$$

where $\sigma=\frac{\kappa-\lambda^2}{\kappa+\lambda^2}$, $z_j=2E_j/\lambda^2$, the energy spectrum is $E_j=\kappa\left(2j+\ell+\frac{3}{2}\right)$ and $j=0,1,2,...$ Moreover, we must choose the scale parameter $\lambda$ such that $\lambda^2<\kappa$. The corresponding discrete weight function reads

$$\omega_j=\left(1-\sigma^2\right)^{\ell+\frac{3}{2}}\frac{\left(\ell+\frac{3}{2}\right)_j\sigma^{2j}}{j!}, \tag{21}$$

The solution of the reference problem, where $\alpha=\hat{\alpha}$ and $\beta=\hat{\beta}$, is obtained as the series solution (5b) that reads

$$\psi_{j,\ell,\kappa}^{(\hat{\alpha},\hat{\beta})}(r)=g_0(E_j)\sum_{n=0}^{\infty}P_n^{(\hat{\alpha},\hat{\beta})}(z_j)\phi_n^{\ell}(\lambda r). \tag{22}$$

Since this solution represents bound states, then the energy weight function $g_0(E_j)$ could be determined by normalization. Using the orthogonality of the basis, $\left\langle\phi_n^{\ell}\middle|\phi_m^{\ell}\right\rangle=\delta_{n,m}$, and the dual orthogonality of the Meixner polynomial that reads $\sum_{n=0}^{\infty}P_n^{(\hat{\alpha},\hat{\beta})}(z_j)P_n^{(\hat{\alpha},\hat{\beta})}(z_k)=\delta_{j,k}/\omega_j$, this normalization gives $g_0(E_j)=\sqrt{\omega_j/\lambda}$.

To compare the bound state wavefunction before and after the change $\{\hat{\alpha},\hat{\beta}\}\mapsto\{\alpha,\beta\}$, we must make sure that we are comparing the same energy level *j*. Now, the change in the initial values of the spectral polynomials will change the energy spectrum. That is, for a given *j*, the energy level $E_j$ at which the wavefunction is evaluated differs before and after the change. Since the original energy spectrum is known and given as $E_j=\kappa\left(2j+\ell+\frac{3}{2}\right)$, we only need to evaluate the new discrete energy spectrum. This could be done by calculating the sorted zeros of the large degree asymptotics ($n\to\infty$) of the spectral polynomial $P_n^{(\alpha,\beta)}(2E/\lambda^2)$. We could also evaluate the sorted energy eigenvalues of a large enough total Hamiltonian matrix $H=H_0+V_{\ell,\kappa}^{(\alpha,\beta)}$. The details of how to construct the matrix elements of the total Hamiltonian for such systems are given below in section 6. For a given $\alpha$ and $\beta$, Table 1 shows the lowest part of the energy spectrum before and after the change $\{\hat{\alpha},\hat{\beta}\}\mapsto\{\alpha,\beta\}$. It is evident that for higher energy levels, the deviation becomes less pronounced. Figure 8 compares the reference solution $\psi_{j,\ell,\kappa}^{(\hat{\alpha},\hat{\beta})}(r)$ (solid trace) to other solutions $\psi_{j,\ell,\kappa}^{(\alpha,\beta)}(r)$ (dashed trace) for the lowest energy bound states $j=0,1,2,3,4,5$. The total potential function for this problem becomes

$$V(r)=\frac{1}{2}\kappa r^2+V_{\ell,\kappa}^{(\alpha,\beta)}(r), \tag{23}$$

with $V_{\ell,\kappa}^{(\hat{\alpha},\hat{\beta})}(r)=0$. In section 7, we derive the potential function $V_{\ell,\kappa}^{(\alpha,\beta)}(r)$ for any given set of parameters $\{\ell,\kappa,\alpha,\beta,\lambda\}$ as shown by Eq. (48).

# 5. The 1D Morse oscillator class

In one dimension with coordinate $x$ and $x_\pm=\pm\infty$, we take the reference Hamiltonian as that of the Morse oscillator. That is, we write

$$H_0=-\frac{1}{2}\frac{d^2}{dx^2}+\frac{\lambda^2}{8}\left(e^{-2\lambda x}+2\mu e^{-\lambda x}\right), \tag{24}$$

where the scale parameter $\lambda$ is a measure of the range of the Morse potential and $\mu$ is a dimensionless real parameter such that if $\mu>-1$ then the energy spectrum is purely continuous. On the other hand, if $\mu<-1$ then the energy spectrum is a mix of continuous and discrete corresponding to scattering states and a finite number of bound states. The following orthonormal basis elements carry a tridiagonal symmetric matrix representation for $H_0$ [i.e., satisfy Eq. (3)]:

$$\phi_n^\nu(\lambda x)=\sqrt{\lambda\Gamma(n+1)/\Gamma(n+\nu)}\,y^{\nu/2}e^{-y/2}L_n^{\nu-1}(y), \tag{25}$$

where $y=e^{-\lambda x}$ and $\nu$ is a positive dimensionless parameter. Using the differential equation, differential property, and recursion relation of the Laguerre polynomials, we can show that Eq. (3) is satisfied with the following coefficients

$$a_n^\nu=(2n+\nu)\left(n+\tfrac{\mu+\nu+1}{2}\right)-n-\tfrac{1}{4}\nu^2, \tag{26a}$$

$$b_n^\nu=-\left(n+\tfrac{\mu+\nu+1}{2}\right)\sqrt{(n+1)(n+\nu)}. \tag{26b}$$

Thus, $\hat{\alpha}=-2\big/(\mu+\nu+1)\sqrt{\nu}$ and $\hat{\beta}=\sqrt{\nu}\left(\frac{\nu/2}{\mu+\nu+1}-1\right)$. The spectral polynomial that satisfies the recursion relation (4) with these recursion coefficients and initial values $P_0(z)=1$, $P_1(z)=\hat{\alpha}z-\hat{\beta}$ is the continuous dual Hahn polynomial whose normalized version reads

$$P_n^{(\hat{\alpha},\hat{\beta})}(z)=\left[\left(\tfrac{\mu+\nu+1}{2}\right)_n\Big/\sqrt{n!(\nu)_n}\right]{}_3F_2\left(\left.\begin{matrix}-n,\frac{\mu+1}{2}+i\sqrt{z},\frac{\mu+1}{2}-i\sqrt{z}\\ \frac{\mu+\nu+1}{2},\frac{\mu+\nu+1}{2}\end{matrix}\right|1\right), \tag{27}$$

where ${}_3F_2\left(\left.\begin{matrix}a,b,c\\ d,e\end{matrix}\right|y\right)=\sum_{m=0}^{\infty}\frac{(a)_m(b)_m(c)_m}{(d)_m(e)_m}\frac{y^m}{m!}$ is the generalized hypergeometric function. The continuous and discrete weight functions associated with these polynomials are

$$\rho(z)=\frac{1}{2\pi}\left|\Gamma\left(\tfrac{\mu+1}{2}+i\sqrt{z}\right)\left[\Gamma\left(\tfrac{\nu}{2}+i\sqrt{z}\right)\right]^2\right|^2\Big/\Gamma(\nu)\left[\Gamma\left(\tfrac{\mu+\nu+1}{2}\right)\right]^2\left|\Gamma\left(2i\sqrt{z}\right)\right|^2, \tag{28a}$$

$$\omega_j = 2\frac{\left[\Gamma\left(\frac{\nu-\mu-1}{2}\right)\right]^2}{\Gamma(\nu)\Gamma(-\mu)}\left(-j-\tfrac{\mu+1}{2}\right)\frac{(\mu-j+2)_j}{j!}\left[\left(\tfrac{\mu+\nu+1}{2}\right)_j\Big/\left(\tfrac{\mu-\nu+3}{2}\right)_j\right]^2, \tag{28b}$$

where $j=0,1,..,N$ and $N$ is the largest integer less than $-(\mu+1)/2$. The bound state energy spectrum is $E_j = -\frac{1}{2}\lambda^2\left(j+\frac{\mu+1}{2}\right)^2$ and we must choose $\nu > -(\mu+1)$. For the full properties of these polynomials, please refer to Appendix B in [7] or section 9.3 of Ref. [6]. The energy functions $f_0(E)$ and $g_0(E_j)$ in the wavefunction expansion are determined by normalization (for the discrete bound states) or asymptotics (for the continuum scattering states). For the latter, we may try the functional form in sections 2 and 3: $f_0(E)\propto\sqrt{\sqrt{z}\rho(z)/\lambda}$. Figure 9 compares the reference series solution $\psi_{\lambda,\mu}^{(\hat\alpha,\hat\beta)}(E,x)$ (solid trace) to the solution $\psi_{\lambda,\mu}^{(\alpha,\beta)}(E,x)$ (dotted trace) for $\alpha\neq\hat\alpha$ and $\beta\neq\hat\beta$.

On the other hand, for pure bound states, the spectral polynomial becomes the finite discrete dual Hahn polynomial whose normalized version reads

$$P_n^{(\hat\alpha,\hat\beta)}(z_j) = \sqrt{\frac{\left(\frac{\mu+\nu+1}{2}\right)_n (N-n+1)_n}{n!(-n-\nu+1)_n}}\;{}_3F_2\left(\left.\begin{matrix}-n,-j,j+\mu+1\\ \frac{\mu+\nu+1}{2},-N\end{matrix}\right|1\right), \tag{29}$$

where $n,j=0,1,2,..,N$. We can obtain $g_0(E_j)$ by normalizing the wavefunction using the orthogonality of the basis, $\left\langle\phi_n^\nu\middle|\phi_m^\nu\right\rangle=\delta_{n,m}$, and dual orthogonality of the dual Hahn polynomial, $\sum_{n=0}^N P_n^{(\hat\alpha,\hat\beta)}(z_j)P_n^{(\hat\alpha,\hat\beta)}(z_k)=\delta_{j,k}/\omega_j$ where (see Eq. (B11) in Appendix B of Ref. [7])

$$\omega_j = (-N-\nu-1)_N\frac{(2j+\mu+1)(-N)_j(N-j+1)_j}{(-N-\nu-1)_j(j+\mu+1)_{N+1}\,j!}, \tag{30}$$

giving $g_0(E_j)=\sqrt{\omega_j/\lambda}$. An alternative is to equate the reference wavefunction series $\psi_{j,\lambda,\mu}^{(\hat\alpha,\hat\beta)}(x)$ to the exact bound state solution of the Morse problem in 1D [8]:

$$g_0(E_j)\sum_{n=0}^{\infty}P_n^{(\hat\alpha,\hat\beta)}(z_j)\phi_n^\nu(\lambda x) = \sqrt{\frac{-\lambda(2j+\mu+1)}{(j!)\Gamma(-j-\mu)}}\,y^{-\frac{\mu+1}{2}}e^{-y/2}Y_j^{\mu/2}(1/y), \tag{31}$$

where $Y_j^\rho(r) = {}_2F_0\left(\left.\begin{matrix}-j,j+2\rho+1\\ —\end{matrix}\right|-r\right)$ is the finite Bessel polynomial with $j=0,1,2,..,N$ and $N$ is the largest integer less than $-\rho-\frac{1}{2}$ (see section 9.13 in [6], Appendix A in [8], or Appendix A in [9]). For a given $\alpha$ and $\beta$, Table 2 gives the whole energy spectrum before and after the change $\{\hat\alpha,\hat\beta\}\mapsto\{\alpha,\beta\}$ whereas Figure 10 compares the reference solution $\psi_{j,\lambda,\mu}^{(\hat\alpha,\hat\beta)}(x)$ (solid-dotted trace) to $\psi_{j,\lambda,\mu}^{(\alpha,\beta)}(x)$ (dashed trace) for the given parameters. In this problem, the total potential function reads

$$V(x) = \frac{\lambda^2}{8}e^{-\lambda x}\left(e^{-\lambda x}+2\mu\right)+V_{\lambda,\mu}^{(\alpha,\beta)}(x), \tag{32}$$

with $V^{(\hat{\alpha},\hat{\beta})}_{\lambda,\mu}(x)=0$. In section 7, we obtain $V^{(\alpha,\beta)}_{\lambda,\mu}(x)$ for any given set of parameters $\{\lambda,\mu,\alpha,\beta,\nu\}$ as shown by Eq. (49).

## 6. Induced bound states and resonances

In this section, we show that if the reference spectral polynomial $P_n^{(\hat{\alpha},\hat{\beta})}(z)$ has a pure continuous spectrum (e.g., the Hermite, Laguerre, Pollaczek, etc.) then it is possible that the change $\{\hat{\alpha},\hat{\beta}\}\mapsto\{\alpha,\beta\}$ may introduce discrete points in the spectrum of $P_n^{(\alpha,\beta)}(z)$ (called "mass points") [10]. This means that if the reference quantum system consists of purely continuous scattering states, then the modified system could pick up new discrete bound states and/or resonances. To demonstrate this curious phenomenon, we investigate the free particle in 3D considered in section 3 where the tridiagonal symmetric Hamiltonian matrix of the reference problem reads

$$\left(H_0\right)_{n,n}=\frac{\lambda^2}{2}a_n^{\ell}=\frac{\lambda^2}{2}\left(2n+\ell+\tfrac{3}{2}\right), \tag{33a}$$

$$\left(H_0\right)_{n,n+1}=\left(H_0\right)_{n+1,n}=\frac{\lambda^2}{2}b_n^{\ell}=\frac{\lambda^2}{2}\sqrt{(n+1)\left(n+\ell+\tfrac{3}{2}\right)}. \tag{33b}$$

Now, the change $\{\hat{\alpha},\hat{\beta}\}\mapsto\{\alpha,\beta\}$ in the initial values of the spectral polynomials will result in an identical total Hamiltonian matrix $H=H_0+V^{(\alpha,\beta)}$ except for the following three elements

$$H_{0,0}=\frac{\lambda^2}{2}\left(\beta/\alpha\right),\qquad H_{0,1}=H_{1,0}=\frac{\lambda^2}{2}\left(1/\alpha\right). \tag{34}$$

This is so, because the change $\{\hat{\alpha},\hat{\beta}\}\mapsto\{\alpha,\beta\}$ is equivalent to $\{a_0,b_0\}\mapsto\{\beta/\alpha,1/\alpha\}$ as could be inferred from the map $P_1^{(\hat{\alpha},\hat{\beta})}(z)=\dfrac{z-a_0}{b_0}\mapsto P_1^{(\alpha,\beta)}(z)=\alpha z-\beta$. See also the discussion in the Appendix leading to the results shown by Eq. (A9) and Eq. (A10). We construct the $N\times N$ tridiagonal symmetric matrix *H* for some large enough integer *N* and obtain its real eigenvalues $\{E_n\}_{n=0}^{N-1}$, which approximate very well the energy spectrum. Equivalently, the energy spectrum (scaled by $2/\lambda^2$) is the union of the set of zeros of the spectral polynomials $P_n^{(\alpha,\beta)}(z)$ in the asymptotic limit $n\to\infty$.

Figure 11 is a snapshot of a video animation* showing $\{E_n\}_{n=0}^{N-1}$ as black dots on the horizontal energy axis and obtained by fixing the parameter $\beta=\hat{\beta}$ but varying $\alpha$ within a certain range of values in the region $\alpha<\hat{\alpha}$. The points on the positive real line corresponds to the continuous spectrum that becomes dense as *N* (the size of the Hamiltonian matrix) gets very large. However, a single energy eigenvalue (circled with red in Figure 11) pops up on the negative real line whose position changes with $\alpha$. It corresponds to a bound state in otherwise continuous energy spectrum. The solid trace in Figure 12 is a plot of this bound state wave

* This video animation is an MP4 file of size 147 KB named "Bound.mp4".

function $\psi_\ell^{(\alpha,\hat{\beta})}(E,r)$ corresponding to the mass point shown in Figure 11 for $\alpha = -0.030$, $E = -34.021$, and $\ell = 1$. The figure also shows the same bound state in a dashed trace but for a large negative value of $\alpha$ that brings it very close to the continuum. Figure 13 is a plot of the energy of this bound state as a function of $\alpha$ for several angular momenta. The black dot on the energy trace corresponds to the critical value of $\alpha$, which we designate as $\alpha_{cr}^{\ell}$, above which no bound state could be induced. Table 3 is a list of values of these critical parameters for the lowest angular momenta.

On the other hand, if we zoom closer into the continuum (e.g., by increasing the value of the scale parameter $\lambda$), we observe the movement of another single energy eigenvalue within the continuum. For a good observation of this phenomenon, the reader may view the video animation[†] of the energy spectrum whose snapshot is shown as Figure 14. This energy eigenvalue, which is shown in the figure as the black dot circled in red, corresponds to a resonance state whose position and width (i.e., life time) changes with $\alpha$. Unfortunately, we were unable to figure out the width of this resonance.

A similar observation could be made if we fix the parameter $\alpha = \hat{\alpha}$ and vary $\beta$ within a certain range of values in the region $\beta < \hat{\beta}$.

# 7. Calculating the induced potential function $V^{(\alpha,\beta)}(x)$

In this section, we evaluate the two-parameter potential function $V^{(\alpha,\beta)}(x)$ induced by the change in the initial values of the spectral polynomials. To accomplish that, we choose a suitable method out of the four given in Section 3 of Ref. [11]. The idea behind those four methods is that if we are given the matrix elements of the potential in a chosen square integrable basis, then we can derive the potential function in configuration space using only the matrix elements of the potential and the basis functions in which they are given. Out of the four methods therein, we found that the first (subsection 3.1 in [11]) and third (subsection 3.3 in [11]) are most suitable for our current problem. This is because the number of non-zero elements of the potential matrix is very small. Out of those two, the third method is our choice due to its higher stability and accuracy. We start by giving a brief outline of the method.

In configuration space with coordinate *x*, we can use Dirac notation to write $\langle x|V|x'\rangle = V(x)\delta(x-x')$ where $\delta(x-x') = \langle x|x'\rangle$. Completeness of the basis, which reads $\sum_{n=0}^{\infty}|\phi_n\rangle\langle\phi_n| = I$ with *I* being the identity, enables us to integrate this equation over $x'$ as follows

$$V(x) = \int\langle x|V|x'\rangle dx' = \sum_{n,m=0}^{\infty}\int\langle x|\phi_n\rangle\langle\phi_n|V|\phi_m\rangle\langle\phi_m|x'\rangle dx' = \sum_{n,m=0}^{\infty}\phi_n(x)V_{n,m}\int\phi_m(x')dx'. \quad (35)$$

The integral on the right side of this equation, $\int\phi_m(x)dx := F_m$, could either be evaluated exactly (if possible) or approximated numerically. For the latter, we choose Gauss quadrature (see, for example, Ref. [12]) associated with the orthogonal polynomial $\{q_n(y)\}$ that appears in the basis elements $\{\phi_n(x)\}$ since all basis elements in this work could be written as $\phi_n(x) = \sqrt{y'(y)\xi(y)}$

[†] This video animation is an MP4 file of size 199 KB named "Resonance.mp4".

$q_n(y)$ where $y = y(x)$, $y'(y) = dy/dx$, and $\xi(y)$ is the associated weight function. That is, $\int_{y_-}^{y_+} \xi(y) q_n(y) q_m(y) dy = \delta_{n,m}$ where $y_\pm = y(x_\pm)$. Therefore, we can write

$$\int \phi_m(x) dx = \int \xi(y) \frac{\phi_m(x(y))}{\xi(y)} \frac{dx}{dy} dy \cong \sum_{k=0}^{K-1} w_k \frac{\phi_m(x(\tau_k))}{\xi(\tau_k) y'(\tau_k)}, \tag{36}$$

for some large enough integer $K$ (called the quadrature order). The abscissa and numerical weights of the quadrature are $\{\tau_k, w_k\}_{k=0}^{K-1}$. Therefore, we can rewrite (36) as follows:

$$F_m = \int \phi_m(x) dx \cong \sum_{k=0}^{K-1} \frac{w_k\, q_m(\tau_k)}{\sqrt{y'(\tau_k)\xi(\tau_k)}} := G_m^K. \tag{37}$$

Consequently, we can write

$$V(x) = \sum_{n,m=0}^{\infty} V_{n,m} F_m\, \phi_n(x) = \langle \phi(x) | V | F \rangle, \tag{38a}$$

$$V(x) \cong \sum_{n,m=0}^{\infty} V_{n,m}\, G_m^K \phi_n(x) = \langle \phi(x) | V | G^K \rangle, \tag{38b}$$

The key advantage of this formula for the potential function $V^{(\alpha,\beta)}$ (as we shall see below) is that there are only three non-zero elements of the potential matrix $V^{(\alpha,\beta)}$. Thus, the infinite sum reduces to only three terms.

Since the total Hamiltonian $H = H_0 + V^{(\alpha,\beta)}$, then we can write $V^{(\alpha,\beta)} = H - H_0$. Now, we have seen in section 6 that the matrix representation of $H$ and $H_0$ in the basis $\{\phi_n(x)\}$ are identical except for three elements. Therefore, the only non-zero elements of the potential matrix are

$$V_{0,0}^{(\alpha,\beta)} = H_{0,0} - (H_0)_{0,0} = \frac{\lambda^2}{2}\left[(\beta/\alpha) - a_0\right], \tag{39}$$

$$V_{0,1}^{(\alpha,\beta)} = H_{0,1} - (H_0)_{0,1} = \frac{\lambda^2}{2}\left[(1/\alpha) - b_0\right] = V_{1,0}^{(\alpha,\beta)}. \tag{40}$$

Substituting these in (38) we obtain the following exact and approximate evaluation of the induced potential function in configuration space

$$V^{(\alpha,\beta)}(x) = V_{0,0}^{(\alpha,\beta)} F_0\, \phi_0(x) + V_{0,1}^{(\alpha,\beta)} \left[F_0\, \phi_1(x) + F_1\, \phi_0(x)\right]. \tag{41a}$$

$$V^{(\alpha,\beta)}(x) \cong V_{0,0}^{(\alpha,\beta)} G_0^K\, \phi_0(x) + V_{0,1}^{(\alpha,\beta)} \left[G_0^K\, \phi_1(x) + G_1^K\, \phi_0(x)\right]. \tag{41b}$$

The approximation in (41b) should improve by increasing the quadrature order $K$.

As illustration, we consider the free particle in 3D discussed in section 3 and section 6 where $a_0 = \ell + \frac{3}{2}$ and $b_0 = \sqrt{\ell + \frac{3}{2}}$. The parameters of this system needed for formulas (41a) and (41b) are as follows:

$$\phi_0(r)=\sqrt{\frac{2\lambda}{\Gamma(\ell+\frac{3}{2})}}\left(\lambda r\right)^{\ell+1}e^{-\lambda^2r^2/2},\qquad \phi_1(r)=\sqrt{\frac{2\lambda}{\Gamma(\ell+\frac{5}{2})}}\left(\lambda r\right)^{\ell+1}e^{-\lambda^2r^2/2}\left(\ell+\tfrac{3}{2}-\lambda^2r^2\right). \tag{42a}$$

$$\xi(y)=\frac{y^{\ell+\frac{1}{2}}e^{-y}}{\Gamma\left(\ell+\frac{3}{2}\right)},\qquad y'(y)=2\lambda\sqrt{y}, \tag{42b}$$

$$q_0(y)=1,\qquad q_1(y)=\frac{\ell+\frac{3}{2}-y}{\sqrt{\ell+\frac{3}{2}}}. \tag{42c}$$

Simple integration yields the following coefficients needed in the exact formula (41a):

$$F_0=\Gamma\left(1+\tfrac{\ell}{2}\right)\sqrt{2^{\ell+1}\big/\lambda\Gamma\left(\ell+\tfrac{3}{2}\right)},\qquad F_1=-F_0\big/2\sqrt{\ell+\tfrac{3}{2}}. \tag{43}$$

Therefore, Eq. (41a) gives the following exact expression for the induced potential function:

$$V_\ell^{(\alpha,\beta)}(r)=\lambda^2\frac{2^{\frac{\ell}{2}}\Gamma\left(\frac{\ell}{2}+1\right)}{\Gamma\left(\ell+\frac{3}{2}\right)}\left(\lambda r\right)^{\ell+1}e^{-\lambda^2r^2/2}\left[\frac{\beta}{\alpha}-\left(\ell+\tfrac{3}{2}\right)+\left(\ell+1-\lambda^2r^2\right)\left(\frac{1/\alpha}{\sqrt{\ell+\frac{3}{2}}}-1\right)\right]. \tag{44}$$

Nonetheless, to obtain a numerical approximation using Gauss quadrature we execute the following procedure. We start by constructing the quadrature matrix $J$ associated with the Laguerre polynomial that appears in the basis (13), which is tridiagonal and symmetric with the following elements

$$J_{n,n}=2n+\ell+\tfrac{3}{2},\qquad J_{n,n+1}=J_{n+1,n}=-\sqrt{(n+1)\left(n+\ell+\tfrac{3}{2}\right)}. \tag{45}$$

Let $\{\tau_k\}_{k=0}^{K-1}$ be the eigenvalues of a $K\times K$ truncated version of this matrix $J$ with $\{\Lambda_{m,k}\}_{m=0}^{K-1}$ being the corresponding normalized eigenvectors. Therefore, the quadrature weights could be written as $w_k=\Lambda_{0,k}^2$.‡ With all ingredients of formula (41b) determined, we can now approximate the potential function $V_\ell^{(\alpha,\beta)}(r)$ for a given set of parameters $\{\ell,\alpha,\beta,\lambda\}$. The result is shown in Figure 15 with solid trace for the exact formula (41a) or (44) and dotted trace for the approximation (41b). The potential well seen in the figure is the reason behind the presence of the single bound state in an otherwise continuous energy spectrum. That bound state corresponds to the mass point with red circle in Figure 11 where $\alpha=-0.030$, $\beta=\hat{\beta}$, $\ell=1$, and $E=-34.021$. Moreover, the potential barrier could account for the resonance that is also observed in the system and shown in Figure 14 but for a different value of $\alpha$.

Figure 16 is a graphical representation of $V^{(\alpha,\beta)}(x)$ associated with the 1D free particle with even symmetry treated in section 2 for a give $\alpha$ and $\beta$:

‡ A more preferred computation of the weights uses only matrix eigenvalues rather than the numerically demanding calculation of eigenvectors. In that computation, we can write $w_k=\prod_{m=0}^{K-2}(\tau_k-\tilde{\tau}_m)\big/\prod_{n\neq k}^{K-1}(\tau_k-\tau_n)$, where $\{\tilde{\tau}_m\}_{m=0}^{K-2}$ are the eigenvalues of a submatrix of $J$ obtained by deleting the top row and left column.

$$V^{(\alpha,\beta)}(x)=\frac{\lambda^2}{\sqrt{2}}e^{-(\lambda x)^2/2}\left[\frac{\beta}{\alpha}-\frac{1}{2}+\left(1+\frac{\sqrt{2}}{\alpha}\right)(\lambda x)^2\right]. \tag{46}$$

Figure 17 is a reproduction of Figure 16 but for odd symmetry where the induced potential becomes

$$V^{(\alpha,\beta)}(x)=\frac{2\lambda^2}{\sqrt{\pi}}(\lambda x)e^{-(\lambda x)^2/2}\left\{\frac{\beta}{\alpha}-\frac{3}{2}+\left(1+\frac{1}{\alpha}\sqrt{\frac{2}{3}}\right)\left[(\lambda x)^2-1\right]\right\}. \tag{47}$$

Figures 18, on the other hand, is a plot of the potential function $V_{\ell,\kappa}^{(\alpha,\beta)}(r)$ for the 3D isotropic oscillator presented in section 4 for a given set of parameters $\{\ell,\kappa,\alpha,\beta,\lambda\}$. This potential function reads as follows

$$\begin{aligned}V_{\ell,\kappa}^{(\alpha,\beta)}(r)&=\sqrt{2^\ell}\,\lambda^2\frac{\Gamma\left(1+\frac{\ell}{2}\right)}{\Gamma\left(\ell+\frac{3}{2}\right)}(\lambda r)^{\ell+1}e^{-(\lambda r)^2/2}\\&\times\left\{\frac{\beta}{\alpha}-\left(\ell+\tfrac{3}{2}\right)\left(\frac{\kappa^2}{\lambda^4}+1\right)+\left(\frac{\kappa^2}{\lambda^4}-1+\frac{1/\alpha}{\sqrt{\ell+\frac{3}{2}}}\right)\left[\ell+1-(\lambda r)^2\right]\right\}\end{aligned} \tag{48}$$

Figures 19 shows the potential function $V_{\lambda,\mu}^{(\alpha,\beta)}(x)$ for the 1D Morse oscillator discussed in section 5 for a given set of parameters $\{\lambda,\mu,\alpha,\beta,\nu\}$:

$$\begin{aligned}V_{\lambda,\mu}^{(\alpha,\beta)}(x)&=\sqrt{2^\nu}\,\frac{\lambda^2}{2}\frac{\Gamma(\nu/2)}{\Gamma(\nu)}y^{\nu/2}e^{-y/2}\\&\times\left\{\frac{\beta}{\alpha}-\frac{\nu}{2}\left(\frac{\nu}{2}+\mu+1\right)+\left[\frac{1/\alpha}{\sqrt{\nu}}+\tfrac{1}{2}(\nu+\mu+1)\right](\nu-y)\right\}\end{aligned} \tag{49}$$

where $y=e^{-\lambda x}$.

# 8. Conclusion

In this work, we started by choosing a complete orthogonal set of functions in configuration space that carry a tridiagonal symmetric matrix representation for a given Hamiltonian operator. The wavefunction of the corresponding system could be written as point-wise convergent series in the chosen basis set. Thus, the expansion coefficients of the series become orthogonal polynomials in the energy that satisfy the resulting three-term recursion relation and contain all physical information about the system. The recursion relation starts with two-parameter initial values forcing the polynomials to change if the values of these two parameters are altered. Consequently, the wavefunction of the system changes with these parameters implying that a two-parameter potential is induced by this change. We gave several examples demonstrating this phenomenon by starting with well-known reference systems that included the free particle in 1D and 3D, the isotropic oscillator in 3D, and the Morse oscillator in 1D. This is not an exhaustive list of reference problems. In fact, we can associate with each exactly solvable potential $V$ (e.g., the Coulomb, Pöschl-Teller, Eckart, etc.) a two-parameter class of exactly solvable quantum systems with its own two-parameter potential $V+V_{\mu}^{(\alpha,\beta)}$. The treatment of these other classes follows the same procedure as in the sample classes investigated in this work. In section 7, we presented a procedure for calculating the two-parameter potential function $V_{\mu}^{(\alpha,\beta)}(x)$.

Additionally, we have assumed in the present work that the modification $\{\hat{\alpha},\hat{\beta}\}\mapsto\{\alpha,\beta\}$ caused the total potential to be in the additive form $V+V_{\mu}^{(\alpha,\beta)}$, which may not be the only possibility. In fact, modification of the reference problem parameter(s) could cause a highly nontrivial deformation of the reference potential *V*. For example, the value of the original reference parameter $\mu$ could change depending on $\alpha$ and $\beta$. Another possibility is that the reference potential *V* gets multiplied by a spatial function $W_{\mu}^{(\alpha,\beta)}$ making the total potential $W_{\mu}^{(\alpha,\beta)}V$ such that $W_{\mu}^{(\hat{\alpha},\hat{\beta})}=1$. Note that we can always write $V+V_{\mu}^{(\alpha,\beta)}=W_{\mu}^{(\alpha,\beta)}V$ with $W_{\mu}^{(\alpha,\beta)}=1+V_{\mu}^{(\alpha,\beta)}/V$. And, so on.

Finally, we have revealed a curious phenomenon where a system with a pure continuous energy spectrum can acquire bound states and/or resonances due to the change $\{\hat{\alpha},\hat{\beta}\}\mapsto\{\alpha,\beta\}$ in the initial values. We demonstrated this phenomenon for a free particle in three dimensions.

**Acknowledgement:** I am grateful to Mourad E. H. Ismail for fruitful and enlightening comments.

## Appendix: Calculation of the continuous weight function

To calculate the continuous component of the weight function $\rho^{(\alpha,\beta)}(z)$ for the spectral polynomials $P_n^{(\alpha,\beta)}(z)$, we use the fact that it is related to the discontinuity of the associated Green's function $G^{(\alpha,\beta)}(z)$ on the positive real line. That is,

$$\rho^{(\alpha,\beta)}(z)=\frac{1}{2\pi\mathrm{i}}\lim_{\varepsilon\to0^{+}}\left[G^{(\alpha,\beta)}(z+\mathrm{i}\varepsilon)-G^{(\alpha,\beta)}(z-\mathrm{i}\varepsilon)\right]=\frac{1}{\pi}\mathrm{Im}\left[G^{(\alpha,\beta)}(z)\right]. \tag{A1}$$

Moreover, the two Green's functions $G^{(\alpha,\beta)}(z)$ and $G^{(\hat{\alpha},\hat{\beta})}(z)$ are related as

$$G^{(\alpha,\beta)}(z)=\frac{G^{(\hat{\alpha},\hat{\beta})}(z)}{1+W(z)G^{(\hat{\alpha},\hat{\beta})}(z)}. \tag{A2}$$

where $W(z)$ is the Wronskian of the two polynomials $P_n^{(\hat{\alpha},\hat{\beta})}(z)$ and $P_n^{(\alpha,\beta)}(z)$ that reads as follows

$$\begin{aligned}W(z)&=b_n\left[P_n^{(\alpha,\beta)}(z)P_{n+1}^{(\hat{\alpha},\hat{\beta})}(z)-P_{n+1}^{(\alpha,\beta)}(z)P_n^{(\hat{\alpha},\hat{\beta})}(z)\right]\\&=b_0\left[P_0^{(\alpha,\beta)}(z)P_1^{(\hat{\alpha},\hat{\beta})}(z)-P_1^{(\alpha,\beta)}(z)P_0^{(\hat{\alpha},\hat{\beta})}(z)\right]=(1-\alpha b_0)z+\beta b_0-a_0\end{aligned} \tag{A3}$$

Therefore, we can relate the two corresponding weight functions by using (A1) in (A2) to obtain

$$\rho^{(\alpha,\beta)}(z)=\frac{\rho^{(\hat{\alpha},\hat{\beta})}(z)}{\left|1+W(z)G^{(\hat{\alpha},\hat{\beta})}(z)\right|^{2}}. \tag{A4}$$

The Green's function of the reference problem, $G^{(\hat{\alpha},\hat{\beta})}(z)$, could be defined formally as the large degree asymptotic of the ratio $Q_n^{(\hat{\alpha},\hat{\beta})}(z)/P_n^{(\hat{\alpha},\hat{\beta})}(z)$. Specifically,

$$G^{(\hat{\alpha},\hat{\beta})}(z)=-\lim_{n\to\infty}\left[Q_n^{(\hat{\alpha},\hat{\beta})}(z)/P_n^{(\hat{\alpha},\hat{\beta})}(z)\right], \tag{A5}$$

where $Q_n^{(\hat{\alpha},\hat{\beta})}(z)$ is a polynomial of degree $n-1$ in $z$ that satisfies the same three-term recursion relation (4a) but with the initial values $Q_0^{(\hat{\alpha},\hat{\beta})}(z)=0$ and $Q_1^{(\hat{\alpha},\hat{\beta})}(z)=1/b_0$. In our calculation, we employed a representation of $G^{(\hat{\alpha},\hat{\beta})}(z)$ in the form of continued fractions with a terminator function $T(z)$:

$$G^{(\hat{\alpha},\hat{\beta})}(z)=\cfrac{-1}{z-a_0-\cfrac{b_0^2}{z-a_1-\cfrac{b_1^2}{z-a_2-.....-\cfrac{b_{N-1}^2}{z-a_N-T(z)}}}}, \tag{A6}$$

for some large enough integer $N$ and where

$$T(z)=\cfrac{b_\infty^2}{z-a_\infty-\cfrac{b_\infty^2}{z-a_\infty-\cfrac{b_\infty^2}{z-a_\infty-....}}}=\frac{b_\infty^2}{z-a_\infty-T(z)}, \tag{A7}$$

where $\{a_\infty,b_\infty\}=\lim_{n\to\infty}\{a_n,b_n\}$ and we have assumed that the limit exists and that it is unique. If the limit does not exist, then we can make a rough numerical approximation by taking $\{a_\infty,b_\infty\}=\{a_N,b_N\}$ for some large enough integer $N$. In any case, solving Eq. (A7) as a quadratic equation in $T(z)$, we obtain

$$T(z)=\frac{z-a_\infty}{2}\pm\frac{1}{2}\sqrt{\left(z-a_\infty+2b_\infty\right)\left(z-a_\infty-2b_\infty\right)}, \tag{A8}$$

which means that the non-zero support of $\rho^{(\hat{\alpha},\hat{\beta})}(z)$ is in the interval $z\in\left[a_\infty-2b_\infty,a_\infty+2b_\infty\right]$ where $T(z)$ becomes a complex function.

If we attempt to rewrite $G^{(\alpha,\beta)}(z)$ in the same continued fractions representation as that of $G^{(\hat{\alpha},\hat{\beta})}(z)$ given by Eq. (A6) while using relation (A2), we arrive at the following alternative relation

$$\left(\alpha b_0\right)G^{(\alpha,\beta)}(z)=G^{(\hat{\alpha},\hat{\beta})}(z)\Big|_{a_0\mapsto\beta/\alpha,\,b_0\mapsto 1/\alpha}. \tag{A9}$$

Consequently, using (A1) in (A9) we obtain the following relation between the two continuous weight functions

$$\left(\alpha b_0\right)\rho^{(\alpha,\beta)}(z)=\rho^{(\hat{\alpha},\hat{\beta})}(z)\Big|_{a_0\mapsto\beta/\alpha,b_0\mapsto 1/\alpha}, \tag{A10}$$

which could have been inferred from the change $P_1^{(\hat{\alpha},\hat{\beta})}(z)=\dfrac{z-a_0}{b_0}\mapsto P_1^{(\alpha,\beta)}(z)=\alpha z-\beta$ except for the factor $\left(1/\alpha b_0\right)$. Formula (A10) could become a good numerical alternative to (A1) when the large degree asymptotics of the recursion coefficients $\{a_n,b_n\}$ grow as $n^\gamma$ with $\gamma>1$.

Finally, and as a simple illustrative example, we consider the Chebyshev class of orthogonal polynomials that satisfy the following three-term recursion relation

$$2xP_n(x)=P_{n-1}(x)+P_{n+1}(x). \tag{A11}$$

Thus, $a_n=0$ and $b_n=1/2$ giving $\hat{\alpha}=2$, $\hat{\beta}=0$ and resulting in the Chebyshev polynomial (of the second kind) $U_n(x)$ with $x_\pm=\pm1$, $G^{(\hat{\alpha},\hat{\beta})}(x)=-2x+2\mathrm{i}\sqrt{1-x^2}$, and $\rho^{(\hat{\alpha},\hat{\beta})}(x)=\frac{2}{\pi}\sqrt{1-x^2}$. The Chebyshev class of polynomials with fixed parity consists of all those that satisfy (A11) and are either odd or even polynomials in $x$. This requires the initial values have $\beta=0$. That is, $P_0^{(\alpha,\beta)}(x)=1$ and $P_1^{(\alpha,\beta)}(x)=\alpha x$. The other well-known Chebyshev polynomial (of the first kind) with fixed parity is $T_n(x)$ that corresponds to $\rho^{(1,0)}(x)=2\big/\pi\sqrt{1-x^2}$. For a general $\alpha$ and $\beta$, on the other hand, the use of the above analysis gives

$$G^{(\alpha,\beta)}(z)=\frac{-2x+2\mathrm{i}\sqrt{1-x^2}}{1+\left[(\alpha-2)x-\beta\right]\left(x-\mathrm{i}\sqrt{1-x^2}\right)}, \tag{A12}$$

$$\rho^{(\alpha,\beta)}(z)=\frac{2\sqrt{1-x^2}\big/\pi}{1+(\alpha x-\beta)[(\alpha-2)x-\beta]}. \tag{A13}$$

Other Chebyshev polynomials without fixed parity correspond to $\alpha=\hat{\alpha}=2$ and $\beta\neq0$. Examples are those of the third and fourth kind with $\alpha=2$ and $\beta=\pm1$ corresponding to the weight functions $\rho^{(2,\pm1)}(x)=\pi^{-1}\sqrt{(1\pm x)/(1\mp x)}$.

## References

[1] T. S. Chihara, *On co-recursive orthogonal polynomials*, Proc. Amer. Math. Soc. **8** (1957) 899

[2] H. A. Slim, *On Co-recursive Orthogonal Polynomials and Their Application to Potential Scattering*, J. Math. Anal. Appl. **136** (1988) 1

[3] G. Szegő, *Orthogonal Polynomials* (American Mathematical Society, 1939)

[4] M. E. H. Ismail, *Classical and Quantum orthogonal polynomials in one variable* (Cambridge University press, 2009)

[5] E. Merzbacher, *Quantum Mechanics*, 3rd ed. (Wiley, 1998)

[6] R. Koekoek, P. A. Lesky and R. F. Swarttouw, *Hypergeometric Orthogonal Polynomials and Their q-Analogues* (Springer, Heidelberg, 2010)

[7] A. D. Alhaidari, *Representation of the quantum mechanical wavefunction by orthogonal polynomials in the energy and physical parameters*, Commun. Theor. Phys. **72** (2020) 015104

[8] A. D. Alhaidari, *Exponentially confining potential well*, Theor. Math. Phys. **206** (2021) 84 [Russian: TMF **206** (2021) 97]

[9] A. D. Alhaidari, *Progressive approximation of bound states by finite series of square-integrable functions*, J. Math. Phys. **63** (2022) 082102

[10] T. S. Chihara, *An Introduction to Orthogonal Polynomials* (Dover, 2011)

[11] A. D. Alhaidari, *Reconstructing the potential function in a formulation of quantum mechanics based on orthogonal polynomials*, Commun. Theor. Phys. **68** (2017) 711

[12] A. D. Alhaidari, *Gauss Quadrature for Integrals and Sums*, Int. J. Pure Appl. Math. Res. **3** (2023) 1

## Tables Caption

**Table 1**: The lower part of the energy spectrum for the 3D isotropic oscillator (in atomic units) before and after the change $\{\hat{\alpha},\hat{\beta}\}\mapsto\{\alpha,\beta\}$. We took the physical parameters: $\ell=1$, $\kappa=3.0$, $\alpha=-\frac{1}{2}$, and $\beta=-3$. The basis scale parameter was set to $\lambda=1.0$.

**Table 2**: The bound states energy spectrum for the 1D Morse oscillator (in atomic units) before and after the change $\{\hat{\alpha},\hat{\beta}\}\mapsto\{\alpha,\beta\}$. We took the physical parameters: $\mu=-10$, $\lambda=0.5$, $\alpha=-1$, and $\beta=4$. The basis parameter was set to $\nu=16$.

**Table 3**: List of the critical values of the parameter $\alpha$ above which no bound state could be induced for the free particle in 3D with the given angular momenta. The parameter $\beta$ is fixed at $\hat{\beta}$ and we took the scale parameter $\lambda=1.0$.

## Figures Caption

**Fig. 1**: The even wavefunction for the free particle in 1D, $\cos(kx)$, shown as solid trace. The series solution of the reference wave equation $\psi^{(\hat{\alpha},\hat{\beta})}(E,x)$ given by Eq. (8) shown as dotted trace, and the wavefunction $\psi^{(\alpha,\beta)}(E,x)$ for $\alpha=-1.0$ and $\beta=-1.2$ shown in dashed trace. We took the energy $E=1.0$ (in atomic units) and $\lambda=1.0$.

**Fig. 2**: The even wavefunction $\psi^{(\alpha,\beta)}(E,x)$ for the free particle in 1D: $\alpha=-2$ and $\beta=-2$ (solid trace), $\alpha=-2$ and $\beta=\hat{\beta}$ (dotted trace), $\alpha=\hat{\alpha}$ and $\beta=-2$ (dashed trace). We took the energy $E=1.0$ (in atomic units) and $\lambda=1.0$.

**Fig. 3**: The weight function $\rho^{(\alpha,\beta)}(z)$ of the spectral polynomials associated with the free particle in 1D with even parity: $\alpha=\hat{\alpha}$ and $\beta=\hat{\beta}$ where $\rho(z)=e^{-z}/\sqrt{\pi z}$ (solid trace), $\alpha=-2$ and $\beta=\hat{\beta}$ (dotted trace), $\alpha=\hat{\alpha}$ and $\beta=-1.2$ (dashed trace), $\alpha=-2$ and $\beta=-1.2$ (dashed-dotted trace).

**Fig. 4**: The odd wavefunction for the free particle in 1D, $\sin(kx)$, shown as solid trace. The series solution of the reference wave equation $\psi^{(\hat{\alpha},\hat{\beta})}(E,x)$ shown as dotted trace, and the wave function $\psi^{(\alpha,\beta)}(E,x)$ for $\alpha=-0.5$ and $\beta=-1.5$ shown in dashed trace. We took the energy $E=1.0$ (in atomic units) and $\lambda=1.0$.

**Fig. 5**: The weight function $\rho^{(\alpha,\beta)}(z)$ of the spectral polynomials associated with the free particle in 1D with odd parity: $\alpha=\hat{\alpha}$ and $\beta=\hat{\beta}$ where $\rho(z)=2\sqrt{z/\pi}\,e^{-z}$ (solid trace), $\alpha=-0.5$ and $\beta=\hat{\beta}$ (dotted trace), $\alpha=\hat{\alpha}$ and $\beta=-1.5$ (dashed trace), $\alpha=-0.5$ and $\beta=-1.5$ (dashed-dotted trace).

**Fig. 6**: The wavefunction for the free particle in 3D, $\sqrt{kr}J_{\ell+\frac{1}{2}}(kr)$, shown as solid trace. The series solution $\psi_{\ell}^{(\hat{\alpha},\hat{\beta})}(E,r)$ given by Eq. (15) shown as dotted trace. The wavefunction

$\psi_\ell^{(\alpha,\beta)}(E,r)$ for $\alpha=1$ and $\beta=2$ (dashed trace), then for $\alpha=2$ and $\beta=1$ (dashed-dotted trace). We took $\ell=1$, the energy $E=2.0$ (in atomic units) and $\lambda=1.0$.

**Fig. 7**: The weight function $\rho^{(\alpha,\beta)}(z)$ of the spectral polynomials associated with the free particle in 3D: $\alpha=\hat{\alpha}$ and $\beta=\hat{\beta}$ where $\rho(z)=z^{\ell+\frac{1}{2}}e^{-z}/\Gamma\left(\ell+\frac{3}{2}\right)$ (solid trace), $\alpha=1$ and $\beta=\hat{\beta}$ (dotted trace), $\alpha=\hat{\alpha}$ and $\beta=2$ (dashed trace), $\alpha=1$ and $\beta=2$ (dashed-dotted trace). We took the angular momentum quantum number $\ell=1$.

**Fig. 8**: Comparison of the reference solution $\psi_{j,\ell,\kappa}^{(\hat{\alpha},\hat{\beta})}(r)$ (solid trace) to other solutions $\psi_{j,\ell,\kappa}^{(\alpha,\beta)}(r)$ (dashed trace) for the lowest energy bound states $j=0,1,2,3,4,5$ of the isotropic oscillator in 3D. We took $\ell=1$, $\kappa=3$ (in atomic units), $\alpha=-\frac{1}{2}$, $\beta=-3$, and the bound states energies are given in Table 1. The scale parameter was set as $\lambda=1.0$.

**Fig. 9**: Comparison of the reference series solution $\psi_{\lambda,\mu}^{(\hat{\alpha},\hat{\beta})}(E,x)$ (solid trace) to the series solution $\psi_{\lambda,\mu}^{(\alpha,\beta)}(E,x)$ (dotted trace) for the scattering states of the Morse oscillator in 1D. We took the physical parameters: $\lambda=0.5$ (in atomic units), $\mu=7$, $\alpha=-1$, $\beta=-2$, and the energy $E=5.0$ (in atomic units). The basis parameter was taken as $\nu=4$.

**Fig. 10**: The exact bound states for the Morse oscillator in 1D, which are given by Eq. (31), are shown in solid traces whereas the series solutions $\psi_{j,\lambda,\mu}^{(\hat{\alpha},\hat{\beta})}(x)$ are shown in dotted traces. The dashed traces are for the series solutions $\psi_{j,\lambda,\mu}^{(\alpha,\beta)}(x)$. We took the physical parameters: $\lambda=0.5$ (in atomic units), $\mu=-10$, $\alpha=-1$, $\beta=4$, and the bound states energies are given in Table 2. The basis parameter was taken as $\nu=16$.

**Fig. 11**: Snapshot of a video animation showing the energy eigenvalues of the total Hamiltonian matrix for the free particle in 3D as black dots on the horizontal energy axis and obtained by fixing the parameter $\beta=\hat{\beta}$ but varying $\alpha$. The snapshot is taken when $\alpha=-0.03$ at which a bound state appears within a red circle at the energy $E\approx-34$. We took $\ell=1$ and the scale parameter $\lambda=1.0$.

**Fig. 12**: The bound state $\psi_\ell^{(\alpha,\hat{\beta})}(E,r)$ that appears in the spectrum of the free particle in 3D with $\ell=1$. The solid trace corresponds to the bound state shown with red circle in Fig. 11 where $\alpha=-0.03$ and $E\approx-34$. The dashed trace corresponds to the same but for a large negative value of $\alpha$ that brings the bound state very close to the continuum where $E\approx-0$.

**Fig. 13**: The energy of the induced bound state for a free particle in 3D as a function of the parameter $\alpha$ with fixed $\beta=\hat{\beta}$ and for several values of the angular momentum. The dot on the energy trace corresponds to the critical value $\alpha=\alpha_{cr}^{\ell}$ above which no bound state could be induced. We took the scale parameter $\lambda=1.0$.

**Fig. 14**: Snapshot of a video animation showing the energy eigenvalues of the total Hamiltonian matrix for the free particle in 3D as black dots on the horizontal energy axis and obtained by fixing the parameter $\beta=\hat{\beta}$ but varying $\alpha$. The snapshot is taken when $\alpha=0.062$ at which a resonance state appears within a red circle at the energy $E\approx69$. We took $\ell=1$ and the scale parameter $\lambda=2.0$.

**Fig. 15**: The induced potential function $V_\ell^{(\alpha,\beta)}(r)$ associated with the free particle in 3D treated in sections 3 and 6 for $\ell=1$, $\alpha=-0.03$, and $\beta=\hat{\beta}$. The exact formula (44) is shown as solid trace while the approximation (41b) is shown as dotted trac. In the calculation, we took the scale parameter $\lambda=1.0$.

**Fig. 16**: The induced potential function $V^{(\alpha,\beta)}(x)$ associated with the free particle in 1D with even symmetry treated in section 2 for $\alpha=-3$ and $\beta=-1$. We took the scale parameter $\lambda=1.0$

**Fig. 17**: A reproduction of Figure 16 but for odd symmetry with $\alpha=-2$ and $\beta=-3$. We took the scale parameter $\lambda=1.0$

**Fig. 18**: The induced potential function $V_{\ell,\kappa}^{(\alpha,\beta)}(r)$ for the 3D isotropic oscillator presented in section 4 with the physical parameters $\left\{\ell=1,\kappa=3,\alpha=-\frac{1}{2},\beta=-3\right\}$. We took the scale parameter $\lambda=1.0$.

**Fig. 19**: The induced potential function $V_{\lambda,\mu}^{(\alpha,\beta)}(x)$ for the Morse oscillator in 1D treated in section 5 with the physical parameters $\left\{\lambda=1,\mu=3,\alpha=-1,\beta=-3\right\}$. The basis parameter $\nu$ was set equal to 2.0.

# Videos Caption:

**Video 1**: Shows the energy eigenvalues (black dots on the horizontal energy axis) of the total Hamiltonian for the free particle in 3D with $\ell=1$. The video is animated by fixing the parameter $\beta=\hat{\beta}$ but varying $\alpha$ within the interval $[0.15,-0.30]$. The energy spectrum consists of a continuous part on the positive real line and a single bound state with energy that moves with $\alpha$ on the negative real line. For the video, we took the scale parameter $\lambda=1.0$. The video is in MP4 format of size 147 KB named "Bound.mp4". It is available upon request from the author.

**Video 2**: Same as Video 1 but the parameter $\alpha$ is varied within the interval $[-0.15,0.20]$. It shows a single resonance energy moving with $\alpha$ within the continuous spectrum. We took the scale parameter $\lambda=2.0$. The video is in MP4 format of size 199 KB named "Resonance.mp4". It is available upon request from the author.

**Table 1**

| $j$ | $E_j(\hat{\alpha},\hat{\beta})$ | $E_j(\alpha,\beta)$ |
|---|---|---|
| 0 | 7.500000 | 2.931422 |
| 1 | 13.500000 | 11.002991 |
| 2 | 19.500000 | 18.210032 |
| 3 | 25.500000 | 24.857519 |
| 4 | 31.500000 | 31.202564 |
| 5 | 37.500000 | 37.373300 |
| 6 | 43.500000 | 43.450120 |
| 7 | 49.500000 | 49.481632 |
| 8 | 55.500000 | 55.493586 |
| 9 | 61.500000 | 61.497851 |
| 10 | 67.500000 | 67.499303 |

**Table 2**

| $j$ | $-E_j(\hat{\alpha},\hat{\beta})$ | $-E_j(\alpha,\beta)$ |
|---|---|---|
| 0 | 2.531250 | 2.014847 |
| 1 | 1.531250 | 1.029680 |
| 2 | 0.781250 | 0.510073 |
| 3 | 0.281250 | 0.393703 |
| 4 | 0.031250 | 0.051556 |

**Table 3**

| $\ell$ | $\alpha_{cr}^{\ell}$ |
|---|---|
| 0 | 0.517974 |
| 1 | 0.252486 |
| 2 | 0.152709 |
| 3 | 0.104756 |

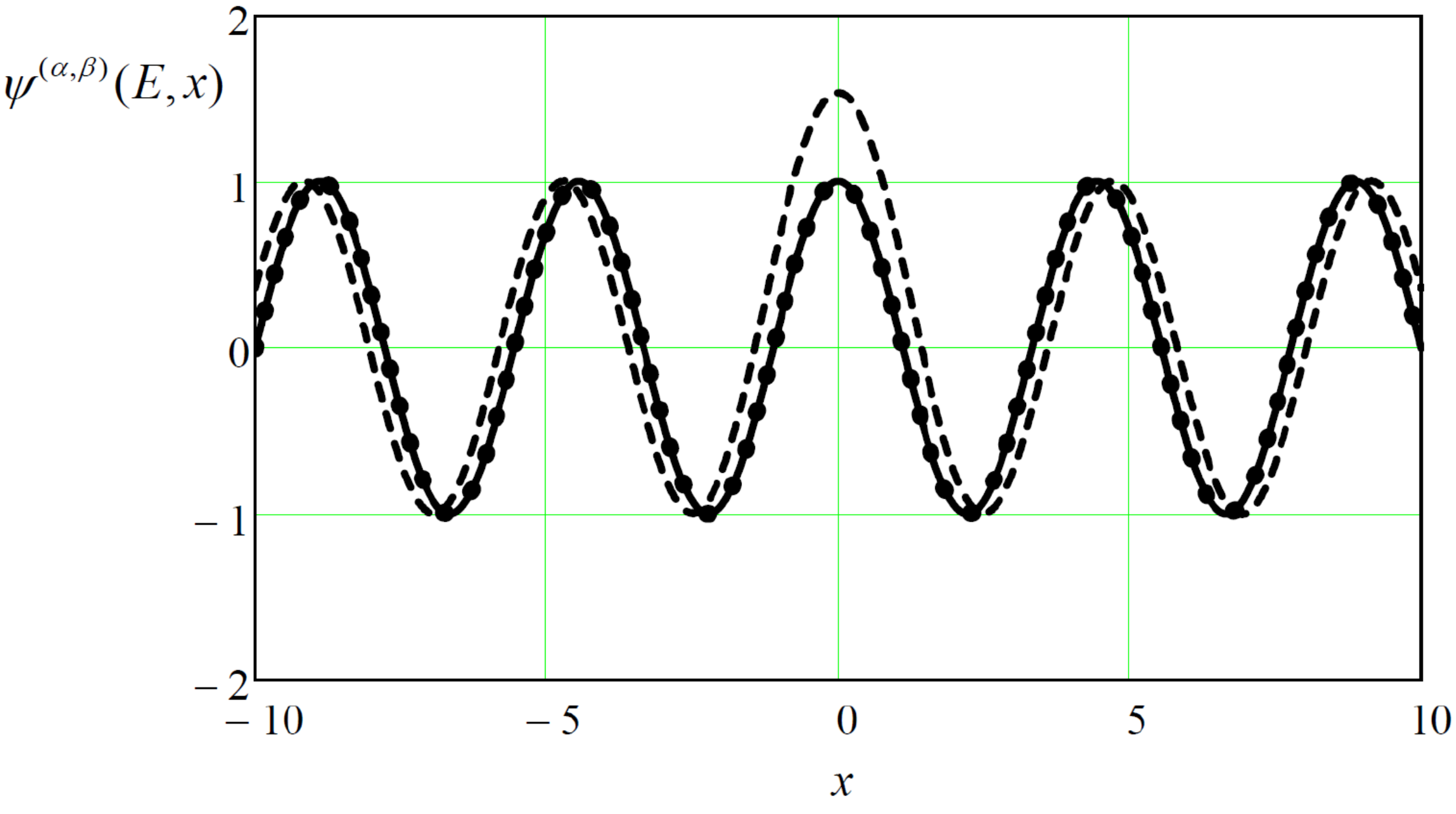
$\psi^{(\alpha,\beta)}(E,x)$
2
1
0
− 1
− 2
− 10
− 5
0
5
10
$x$

**Fig. 1**

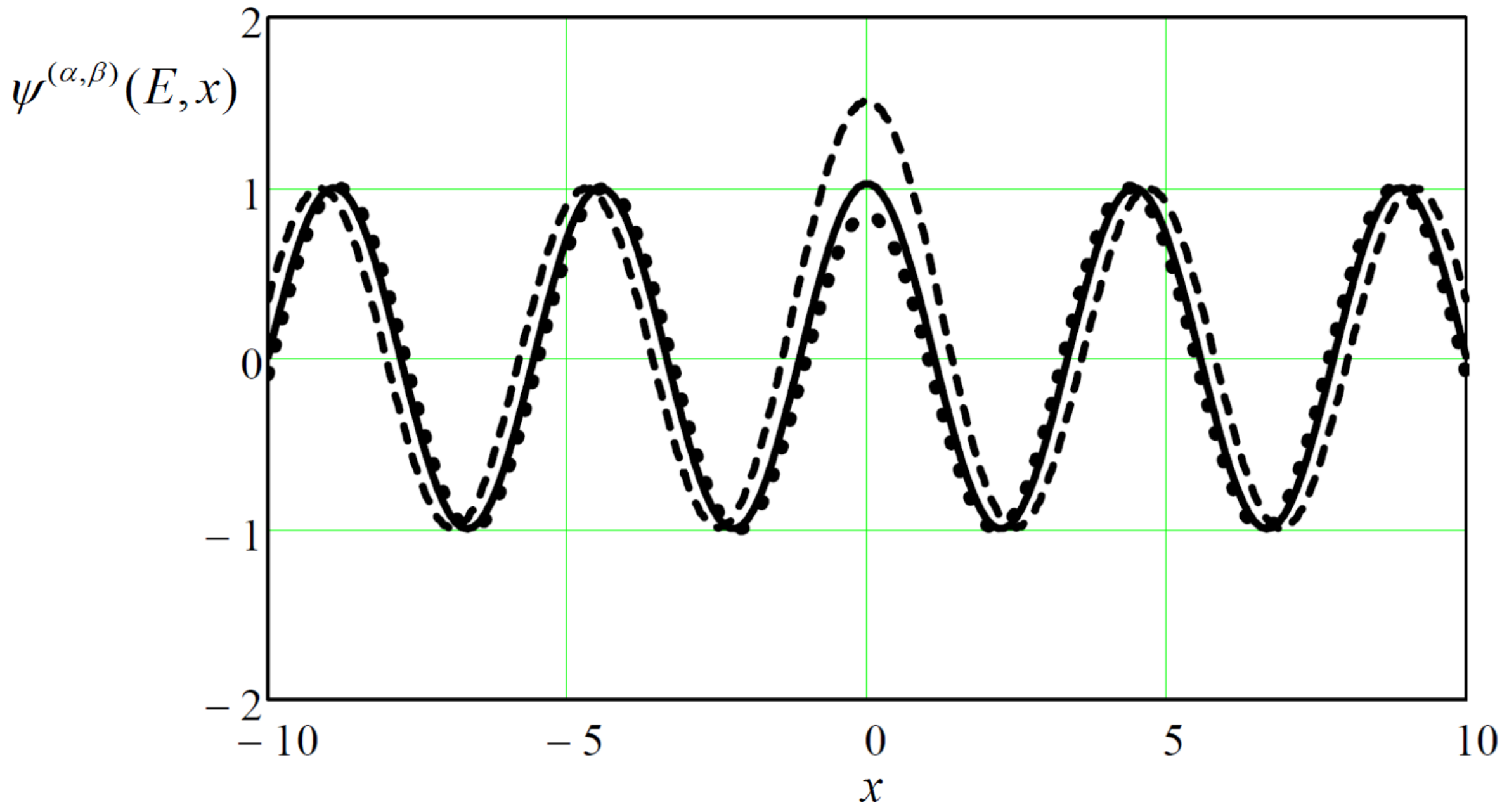
$\psi^{(\alpha,\beta)}(E,x)$
2
1
0
− 1
− 2
− 10
− 5
0
5
10
$x$

**Fig. 2**

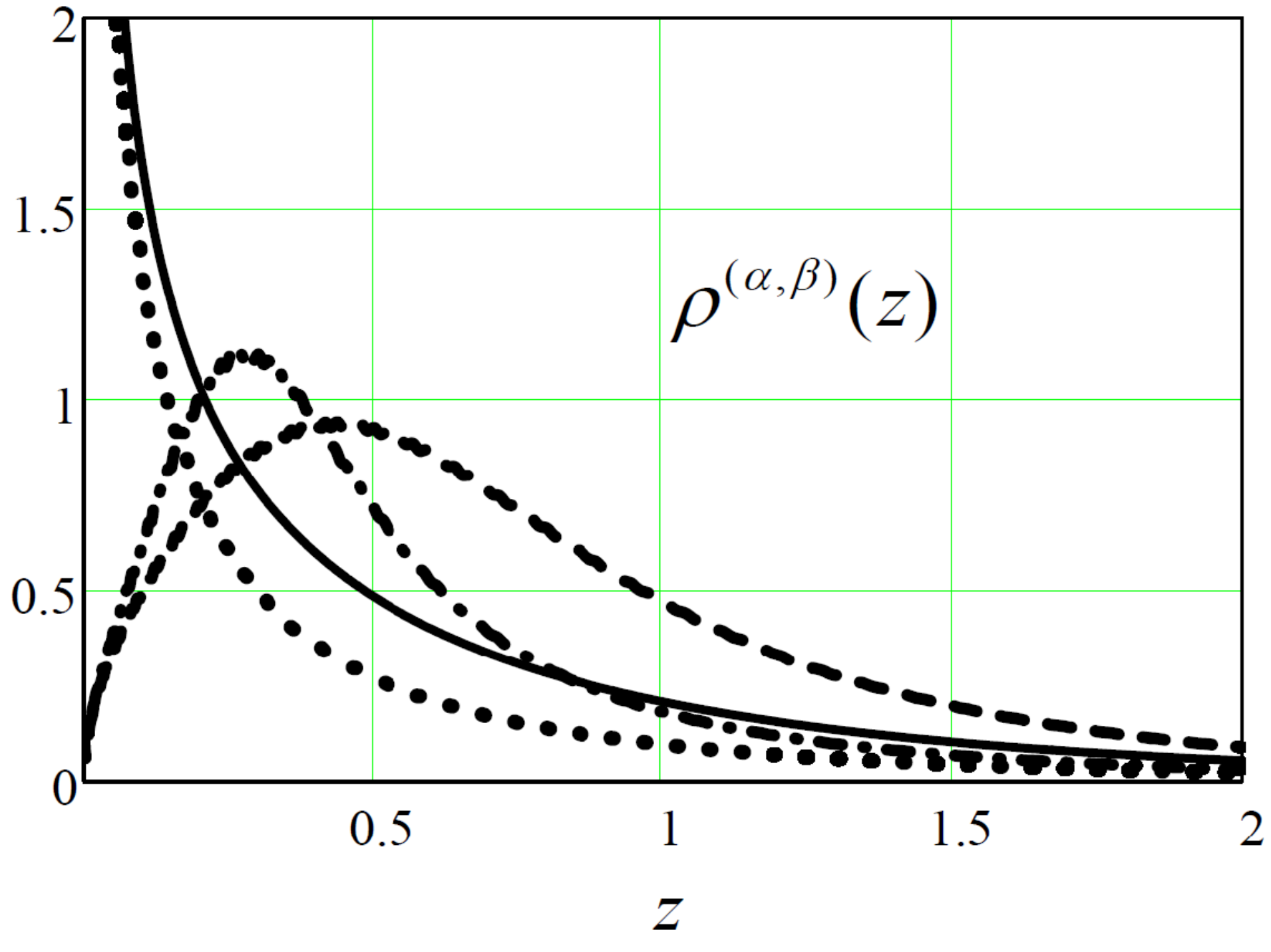

2
1.5
1
0.5
0
$\rho^{(\alpha,\beta)}(z)$
0.5
1
1.5
2
$z$

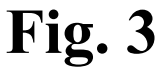

**Fig. 3**

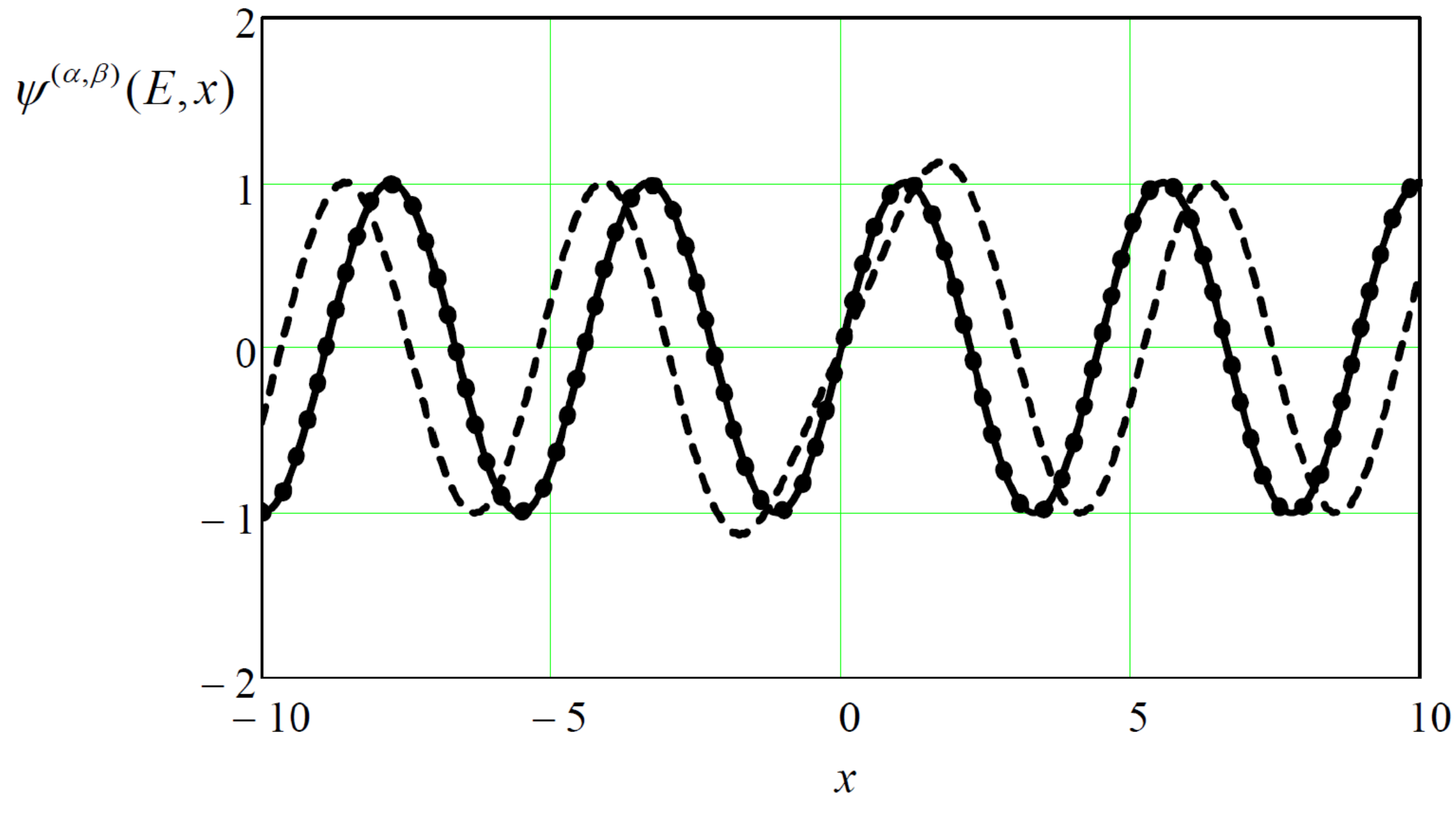

$\psi^{(\alpha,\beta)}(E,x)$
2
1
0
− 1
− 2
− 10
− 5
0
5
10
$x$

**Fig. 4**

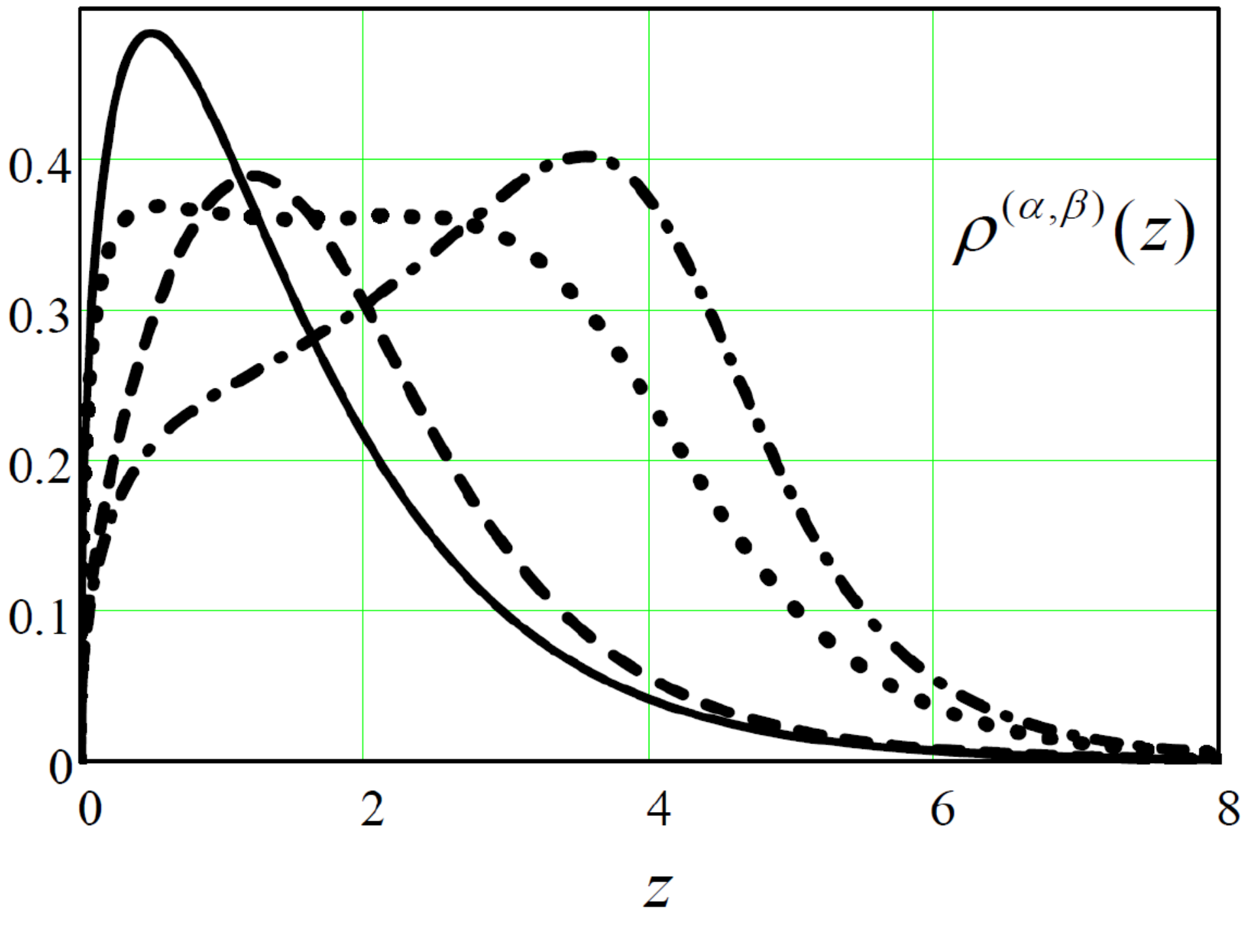

0.4
0.3
0.2
0.1
0
0
2
4
6
8
$z$
$\rho^{(\alpha,\beta)}(z)$

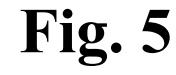

**Fig. 5**

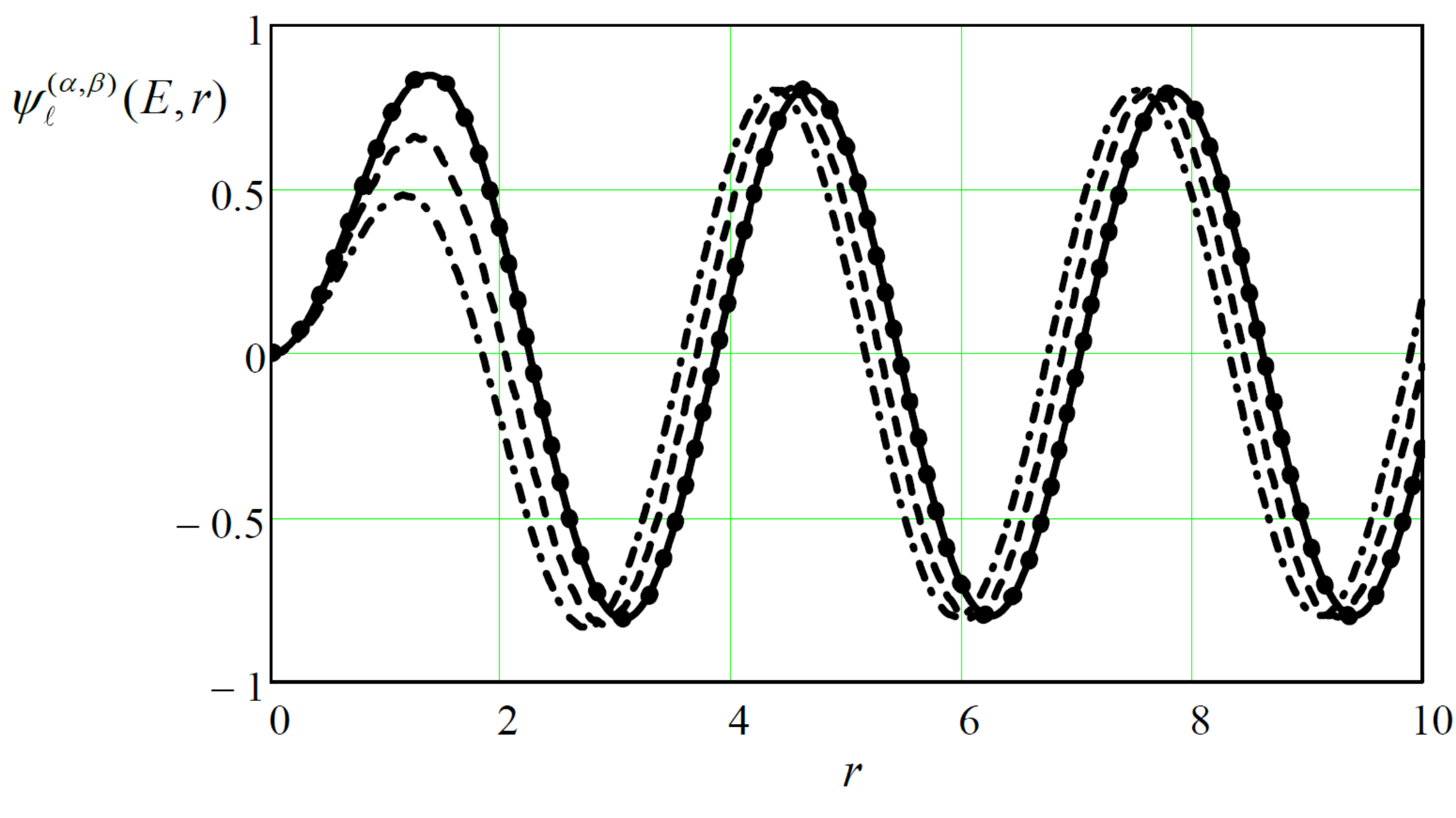

$\psi_\ell^{(\alpha,\beta)}(E,r)$
1
0.5
0
− 0.5
− 1
0
2
4
6
8
10
$r$

**Fig. 6**

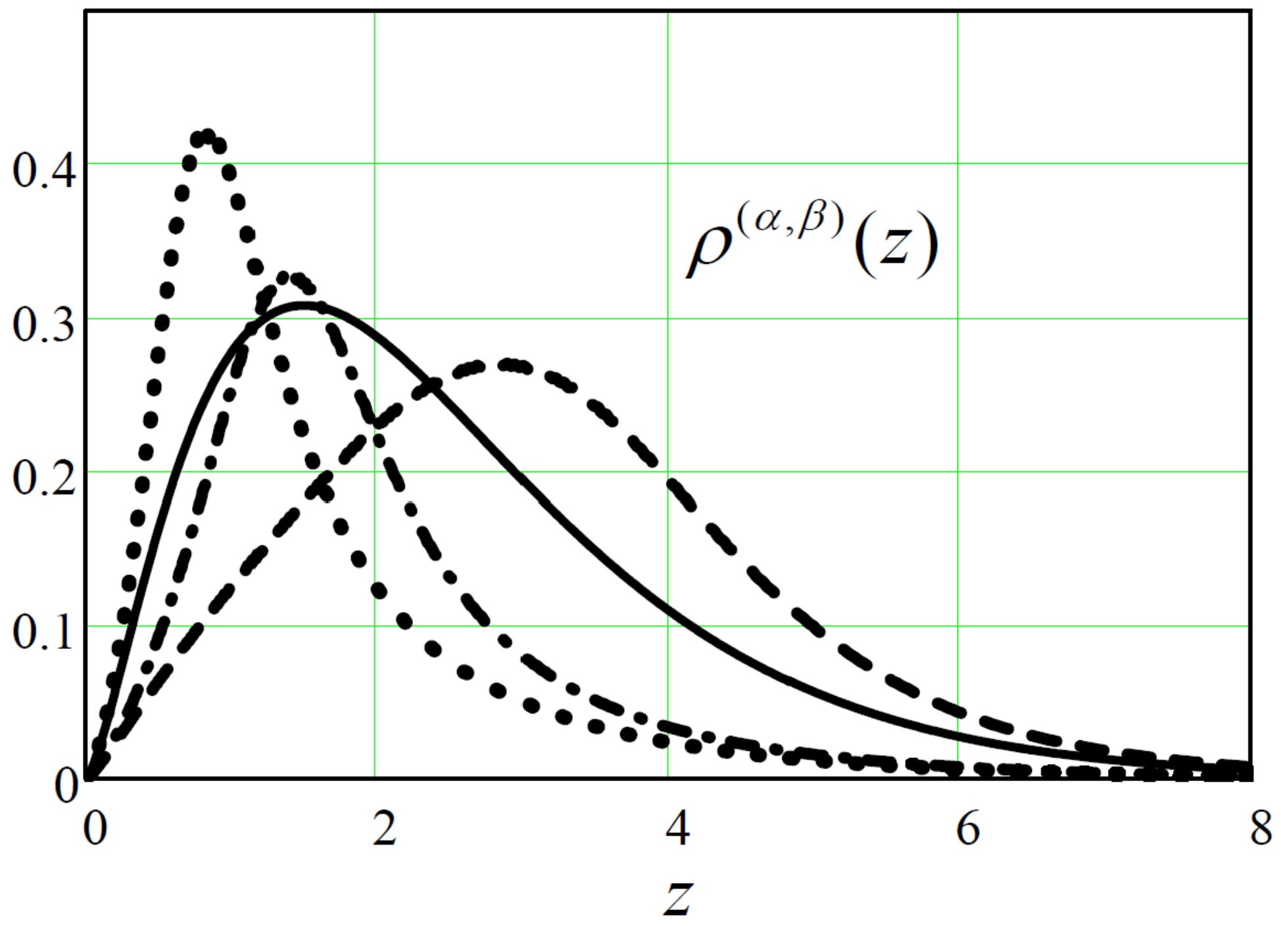

**Fig. 7**

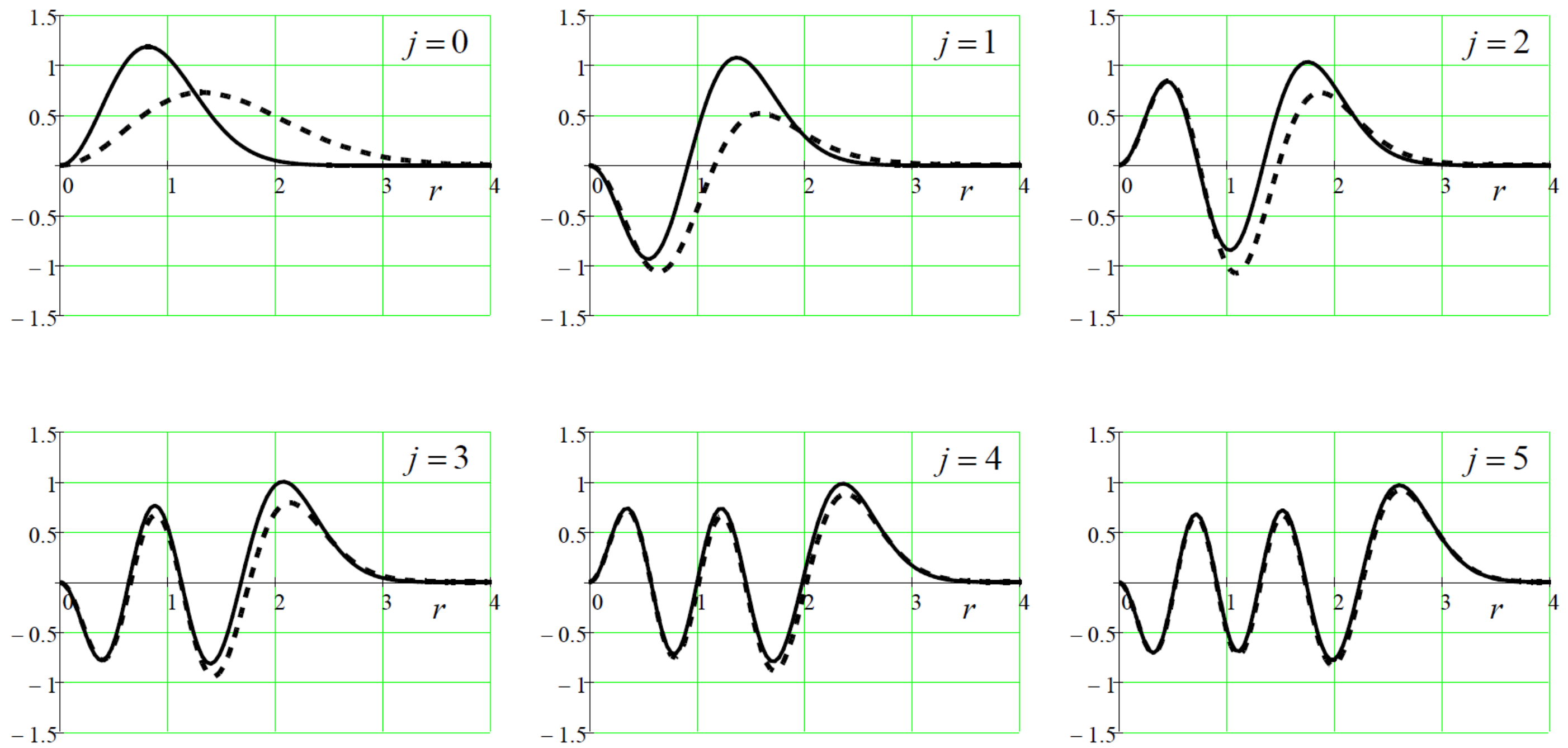

**Fig. 8**

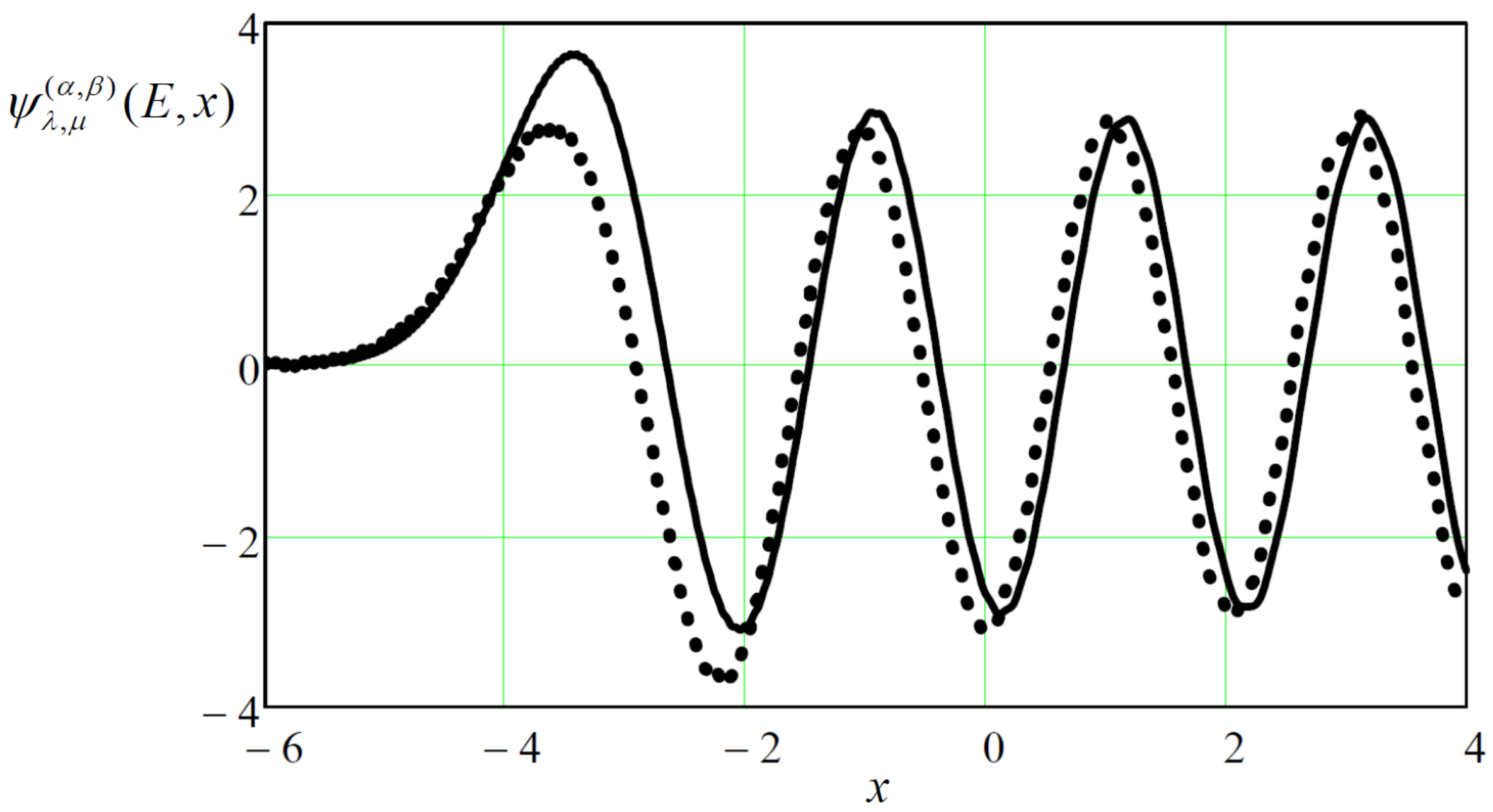

**Fig. 9**

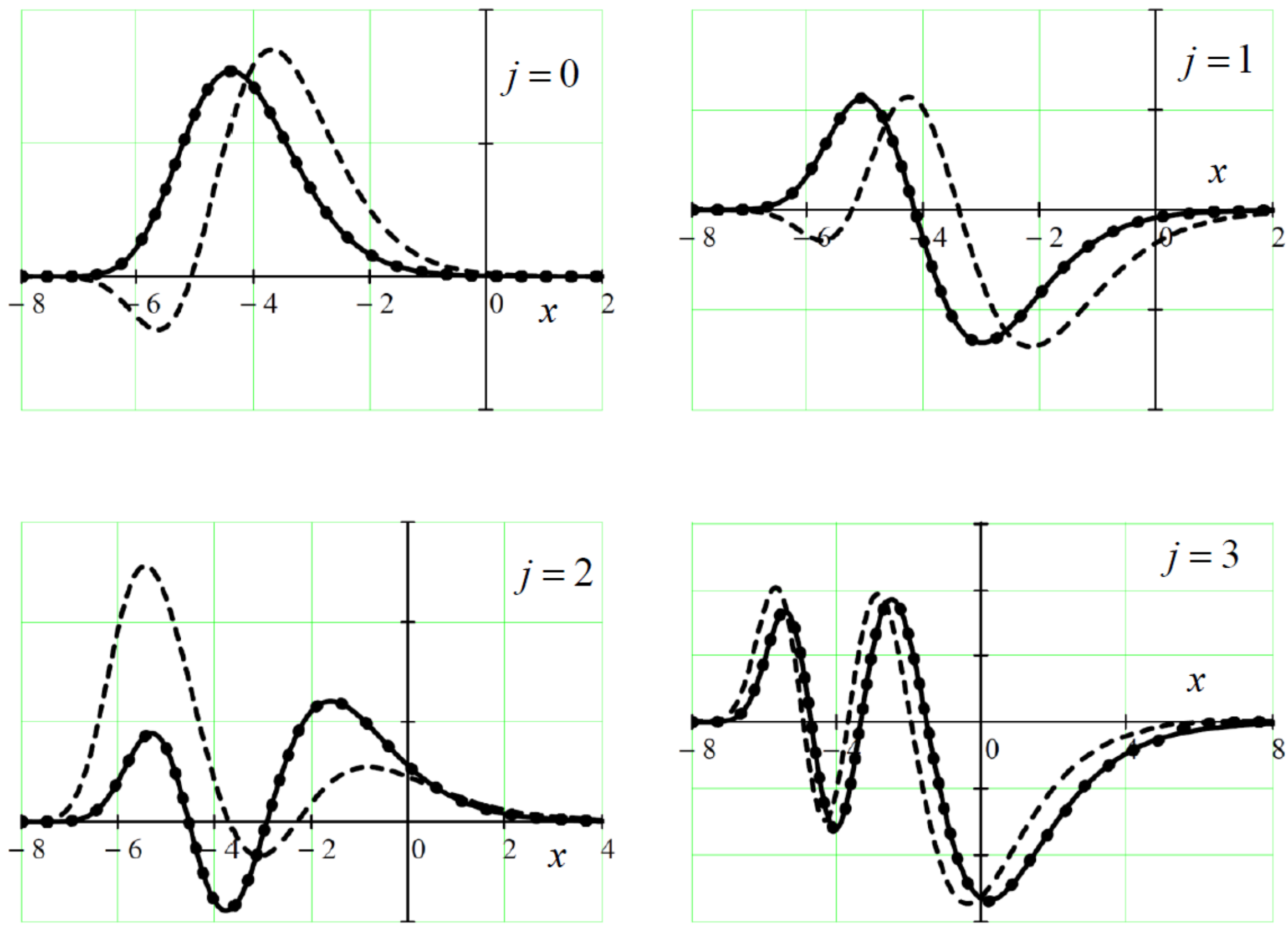

**Fig. 10**

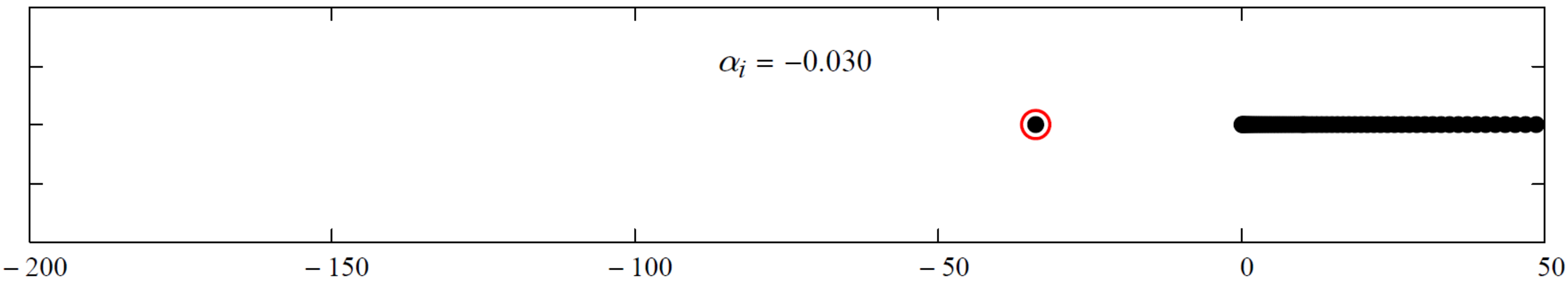

$\alpha_i = -0.030$
– 200
– 150
– 100
– 50
0
50

**Fig. 11**

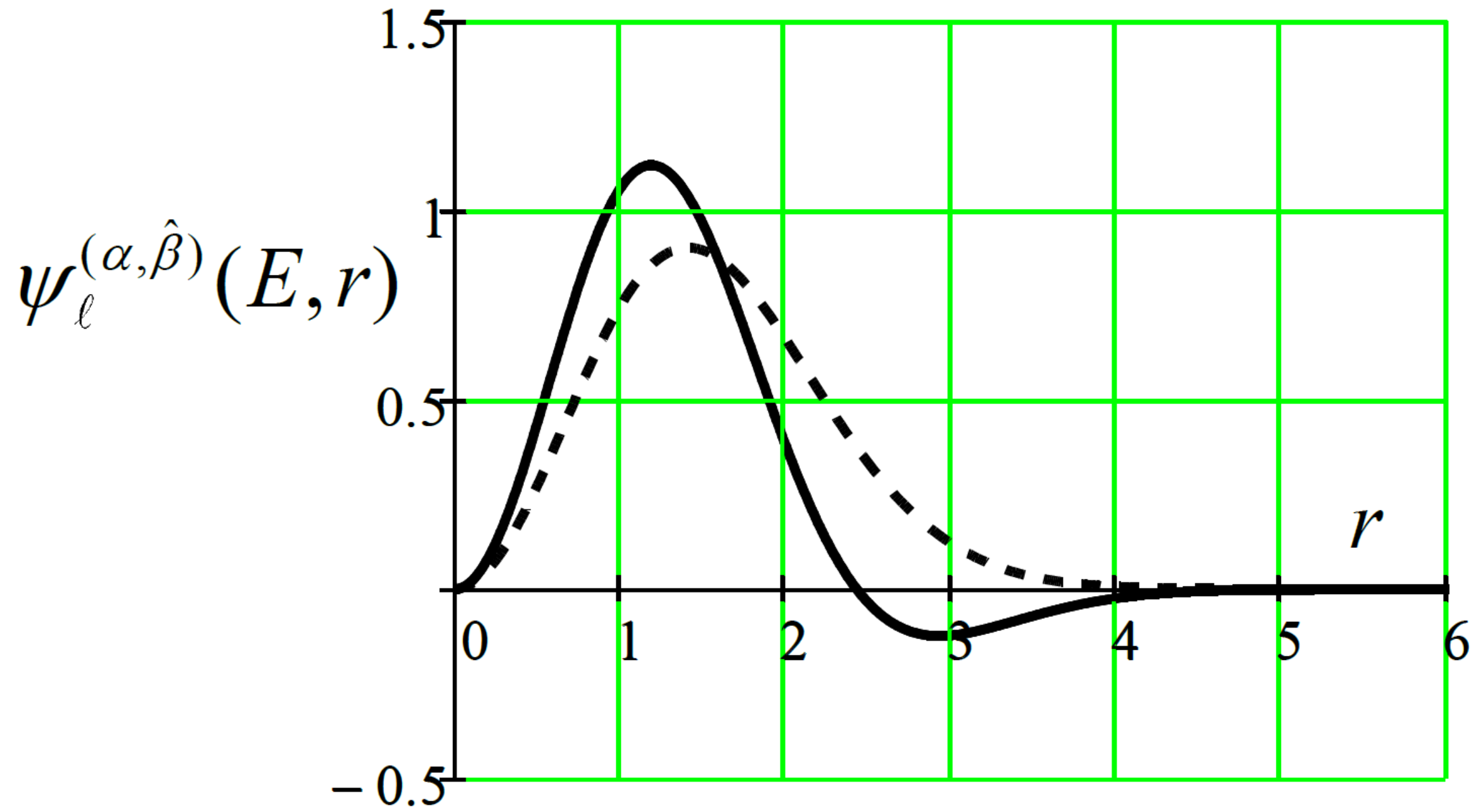

1.5
1
0.5
– 0.5
$\psi_{\ell}^{(\alpha,\hat{\beta})}(E,r)$
$r$
0
1
2
3
4
5
6

**Fig. 12**

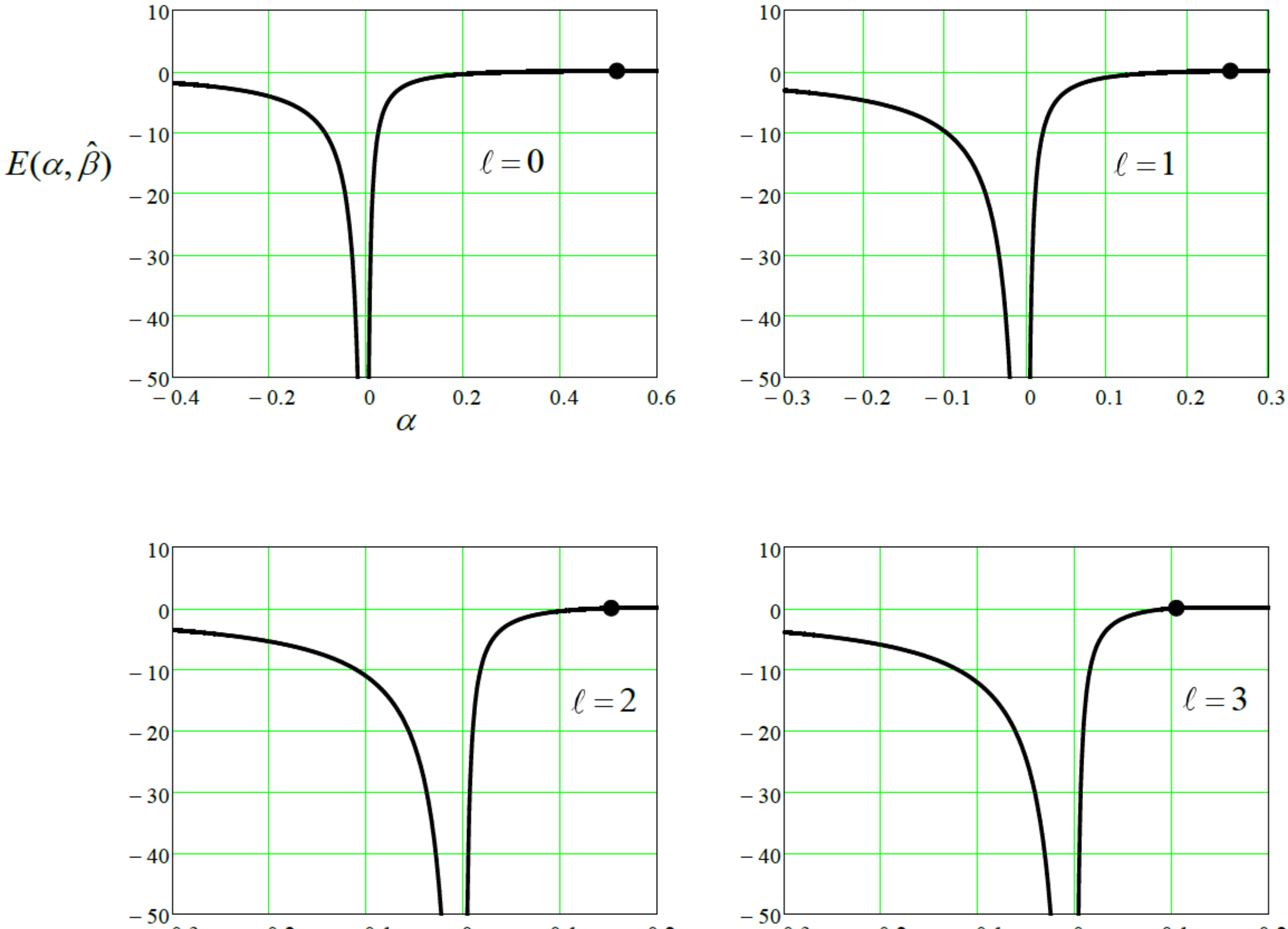

**Fig. 13**

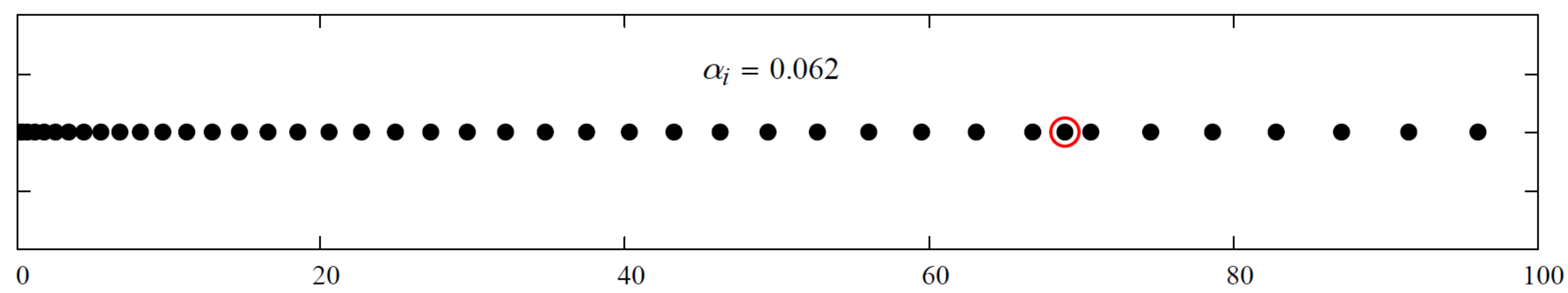

**Fig. 14**

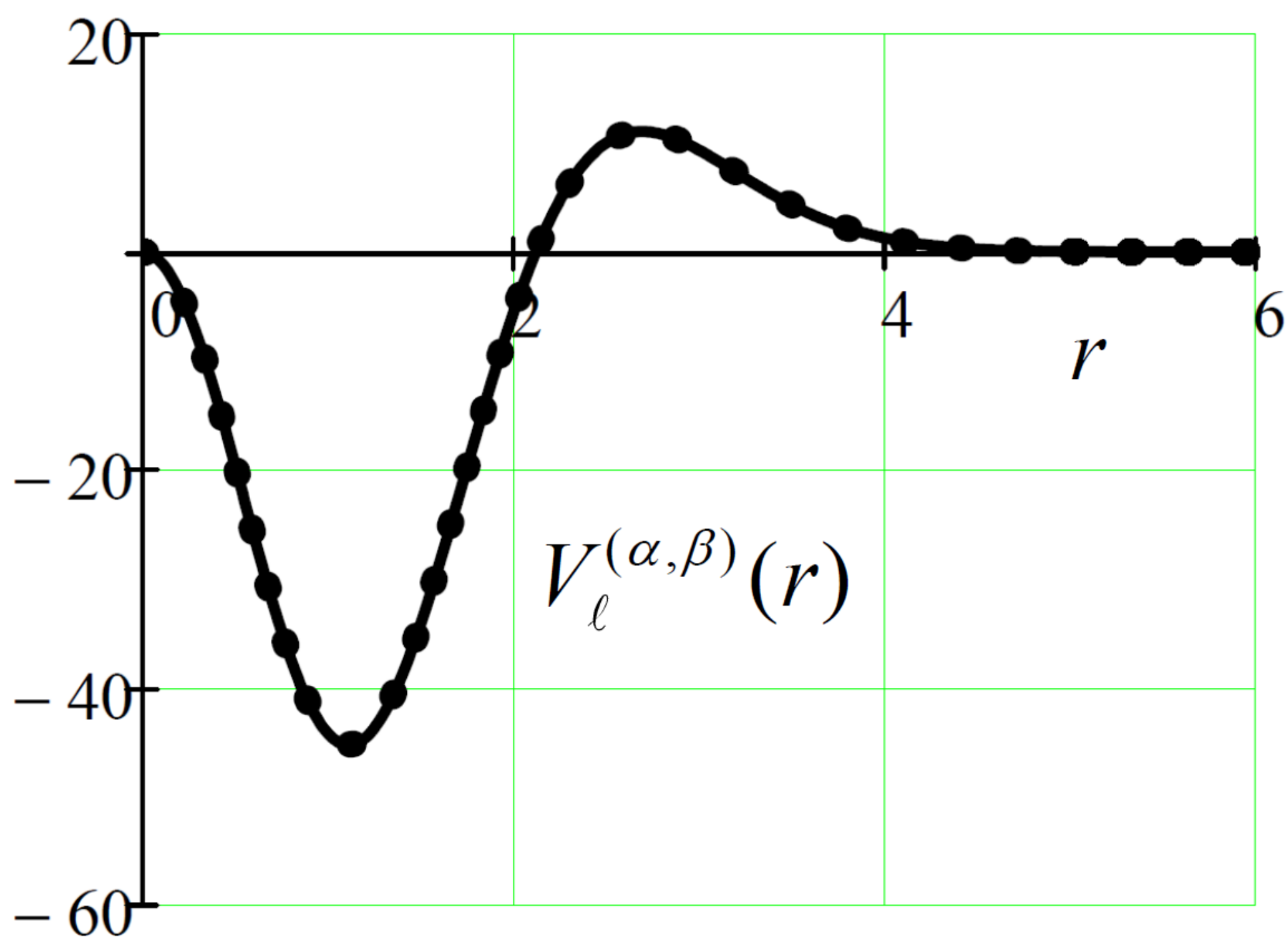

**Fig. 15**

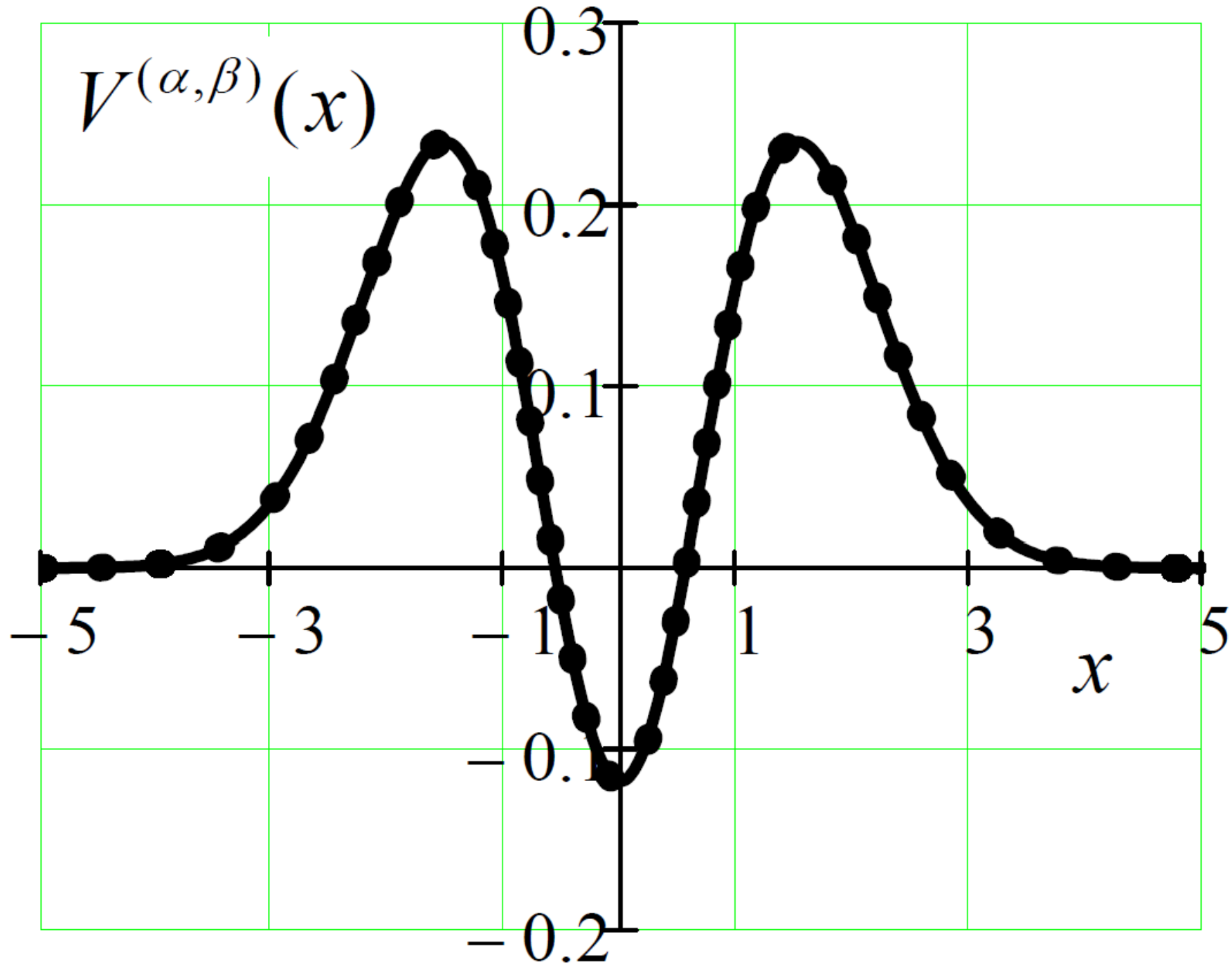
$V^{(\alpha,\beta)}(x)$
0.3
0.2
0.1
– 0.1
– 0.2
– 5
– 3
– 1
1
3
5
$x$

**Fig. 16**

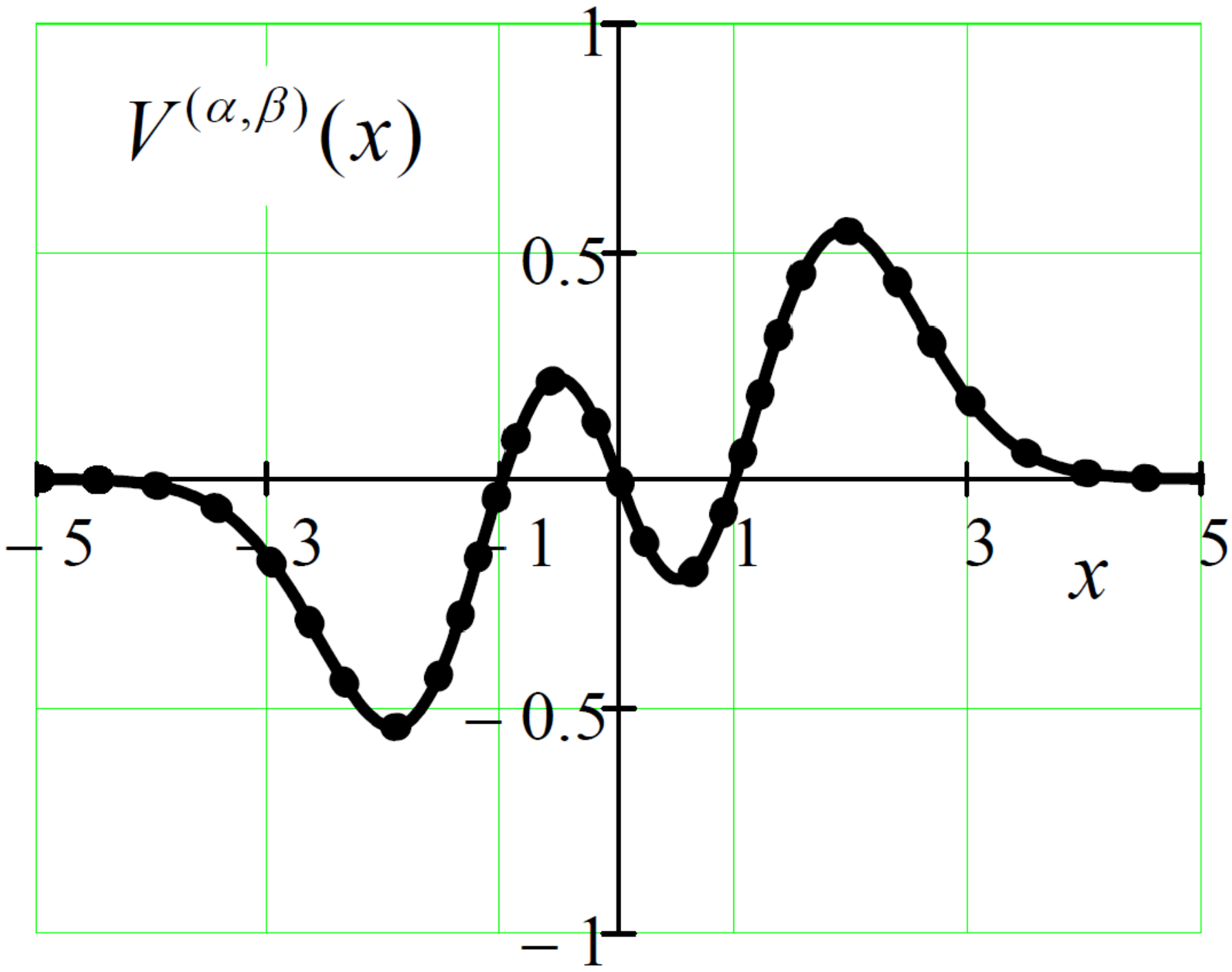
$V^{(\alpha,\beta)}(x)$
1
0.5
– 0.5
– 1
– 5
– 3
– 1
1
3
5
$x$

**Fig. 17**

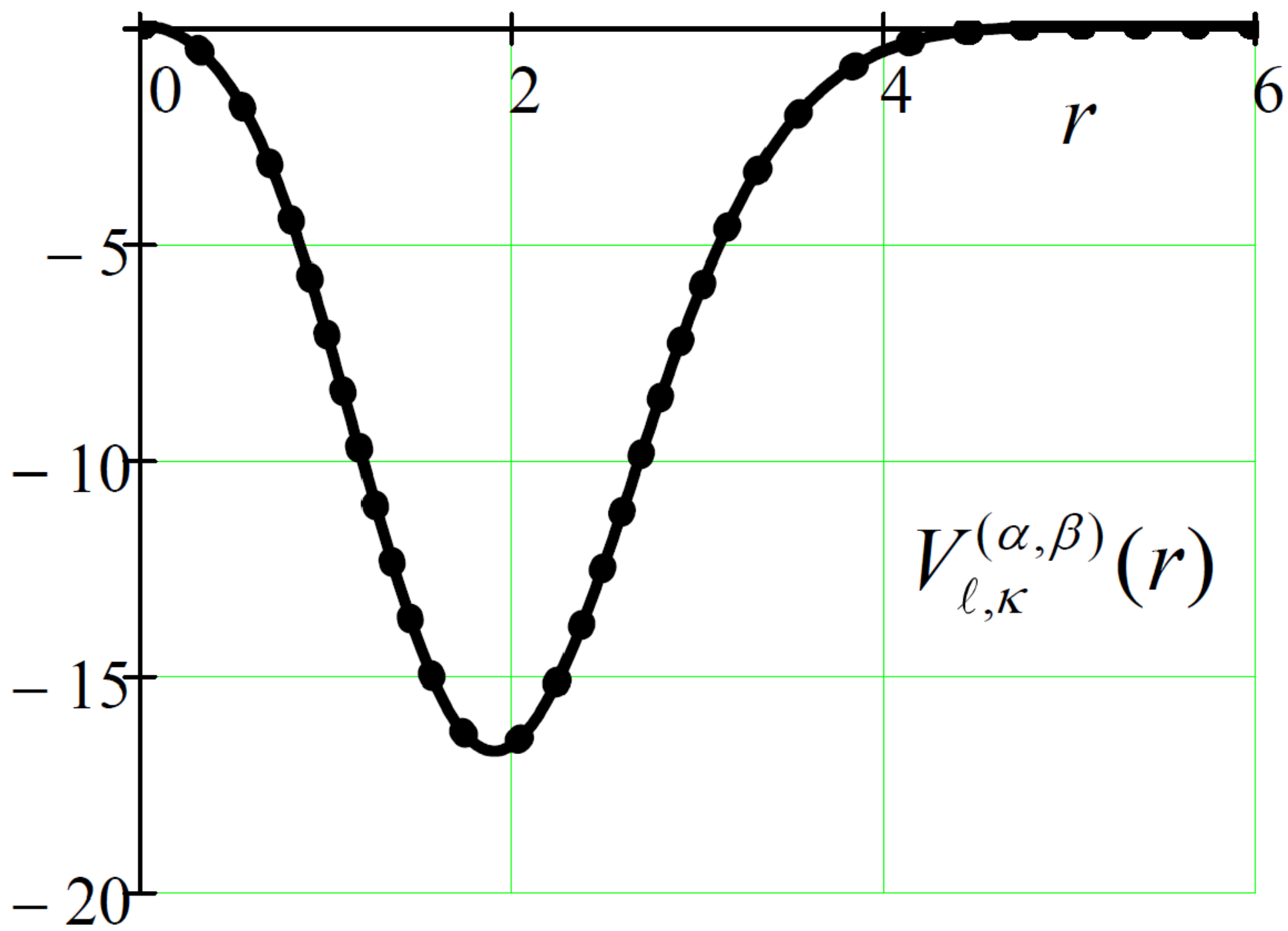
0
2
4
6
r
– 5
– 10
– 15
– 20
$V_{\ell,\kappa}^{(\alpha,\beta)}(r)$

**Fig. 18**

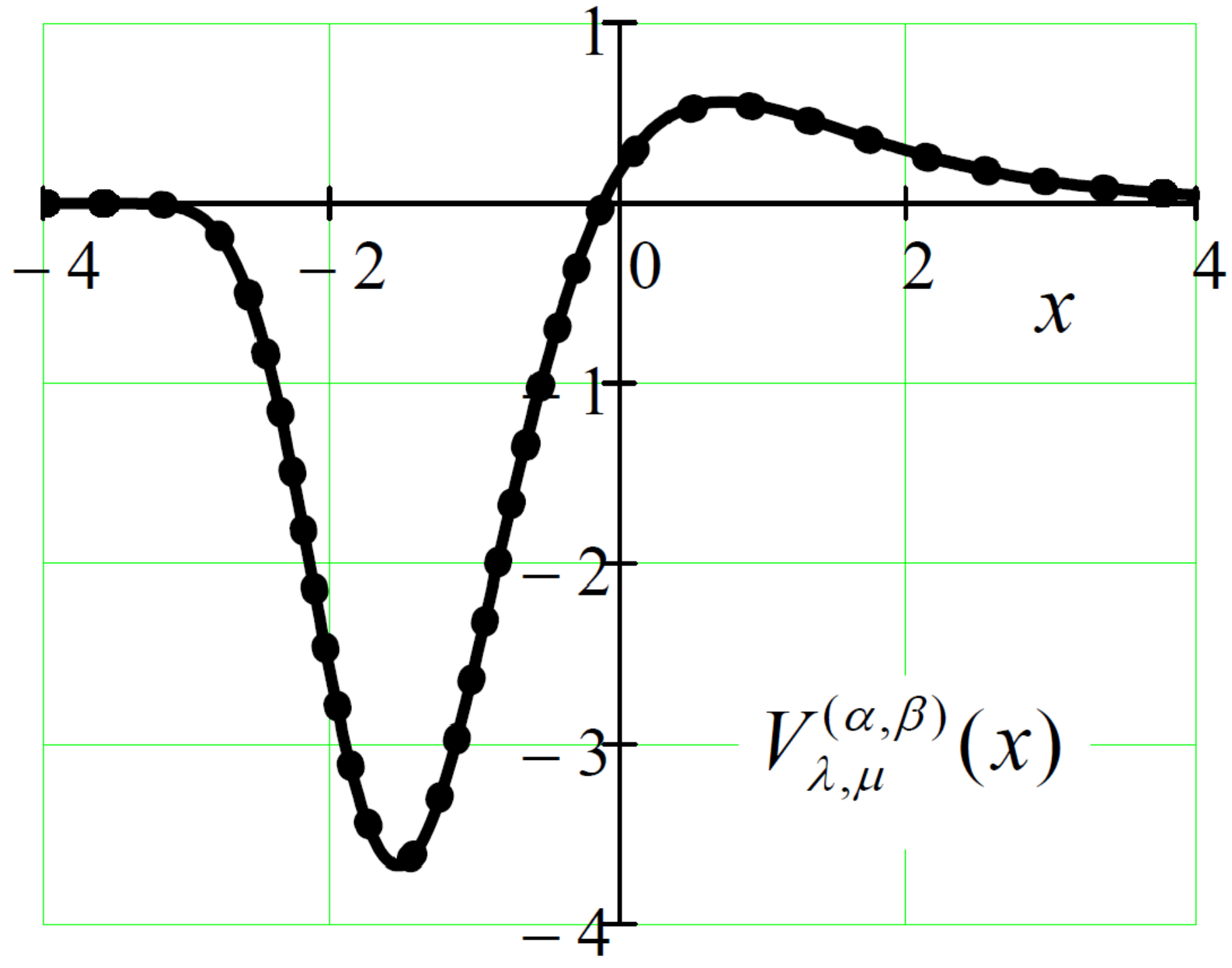
1
– 4
– 2
0
2
4
x
– 1
– 2
– 3
– 4
$V_{\lambda,\mu}^{(\alpha,\beta)}(x)$

**Fig. 19**